\documentclass[preprint,prd,aps,showpacs,showkeys,nofootinbib]{revtex4}
\usepackage{amsmath}
\usepackage{graphicx}
\usepackage{float}
\usepackage[backref]{hyperref}
\usepackage{amstext}
\usepackage{CJKutf8}
\begin{document}


\title{Lepton-flavor violation in the left-right supersymmetric  standard  model}

\author{Yong-Kang Huang$^1$\footnote{2207301041@st.gxu.edu.cn}, Jin-Lei Yang$^{2,3}$\footnote{jlyang@hbu.edu.cn}, Sheng-Kai Cui$^1$, Tai-Fu Feng$^{1,2,3,4}$\footnote{fengtf@hbu.edu.cn}
	}

\affiliation{$^1$Department of Physics, Guangxi University, Nanning, 530004, China\\
	$^2$Department of Physics, Hebei University, Baoding, 071002, China\\
	$^3$Key Laboratory of High-precision Computation and Application of
	Quantum Field Theory of Hebei Province, Baoding, 071002, China\\
	$^4$Department of Physics, Chongqing University, Chongqing 401331, China}

\begin{abstract}
 In the Standard Model (SM), the charged lepton-flavor violation (CLFV) processes are forbidden, so the observation of  CLFV  represents a clear signal of new physics that goes beyond the Standard Model. In this work, we focus on the CLFV processes $l_{j}^{-}\to l_{i}^{-}\gamma $  and $l_{j}^{-}\to l_{i}^{-}l_{i}^{-}l_{i}^{+}$ in LRSSM. Considering the constraints of the updated experimental datas, the numerical results show that the new contributions of $SU{{\left( 2 \right)}_{R}}$ gauge interaction and a large number of other new particles, such as ${{Z}^{'}}$ boson, double-charged Higgs-bosons  in LRSSM can make significant contributions to the CLFV processes $l_{j}^{-}\to l_{i}^{-}\gamma $ and $l_{j}^{-}\to l_{i}^{-}l_{i}^{-}l_{i}^{+}$, which is much different from the ones predicted in other SUSY models. This work provide a theoretical base for finding the LFV phenomena in new physics.

\end{abstract}

\maketitle

\section{Introduction\label{sec1}}

Lepton-flavor violation (LFV) is currently a prominent physical phenomenon that has undergone extensive theoretical and experimental investigation. Experimental observations, particularly in neutrino oscillations, indicate the existence of LFV. However, within the Standard Model (SM), lepton number conservation and strictly zero neutrino masses contradict these oscillation phenomena, highlighting the imperfections of the SM theory. Consequently, there is a necessity to expand beyond the SM framework. Various new physical theories beyond the SM allow for the possibility of LFV, making it a crucial tool for testing these theories. Recent experimental observations have established upper limits for the branching ratios of several physical processes, as detailed in TABLE \ref{tab1}.

If the lepton-flavor violation  phenomenon is observed in the future, it will provide strong support for new physics beyond the SM. Among the various possibilities for expanding the SM with new physics, the supersymmetry (SUSY) theory is a popular and promising direction. One particular supersymmetric model called the Left-Right Supersymmetric Standard Model (LRSSM) has garnered significant attention due to its numerous advantages and favorable properties. Left-right supersymmetric models have been explored before \cite{Mohapatra:1974hk,Senjanovic:1975rk,Huitu:1997rr,Huitu:1999tg}.
 While the LFV phenomenon has been extensively studied in other SUSY models, its exploration in the LRSSM remains limited \cite{Frank:2004hg7,PhysRevD.70.036004_8}. In this study, we aimed to investigate the LFV phenomenon in the LRSSM by analysing and calculating processes $l_{j}^{-}\to l_{i}^{-}\gamma $ and $l_{j}^{-}\to l_{i}^{-}l_{i}^{-}l_{i}^{+}$. We further conducted a comprehensive numerical analysis to compare our results with those of the Minimal Supersymmetric Standard Model (MSSM).

In LRSSM, the $SU{{\left( 2 \right)}_{R}}$ gauge interaction provides additional vertices and makes  additional contributions to the branching ratios of $l_{j}^{-}\to l_{i}^{-}\gamma $ and $l_{j}^{-}\to l_{i}^{-}l_{i}^{-}l_{i}^{+}$ processes. The large number of Higgs bidoublets and triplets,  double-charged Higgs-bosons in LRSSM all contribute to $l_{j}^{-}\to l_{i}^{-}\gamma$ and $l_{j}^{-}\to l_{i}^{-}l_{i}^{-}l_{i}^{+}$  processes, which is quiet different from the case in other models (e.g., the tree-level contribution of $l_{j}^{-}\to l_{i}^{-}l_{i}^{-}l_{i}^{+}$ processes). The paper is organized as follows: Part II provides an explanation of the basic theoretical structure of the LRSSM, including the mass matrix and necessary theoretical derivations. In Part III, we conduct a detailed analysis of the process of calculating $l_{j}^{-}\to l_{i}^{-}\gamma $ and $l_{j}^{-}\to l_{i}^{-}l_{i}^{-}l_{i}^{+}$, and derive formulas for branching ratios. Part IV presents a numerical analysis of the two processes, where we select parameters and create graphs illustrating the changes in branching ratios with relevant sensitive parameters. Finally, in Part V, we summarize our work and draw conclusions.

\begin{table*}
	\begin{tabular*}{\textwidth}{@{\extracolsep{\fill}}lll@{}}
		\hline
		LFV process & Present limit & Future sensitivity\\
		\hline
		$\mu\rightarrow e\gamma$ & $<4.2\times10^{-13}$ & $\sim6\times10^{-14}$ \cite{baldini2013meg1}\\
		$\mu\rightarrow 3e$ & $<1\times10^{-12}$ \cite{BELLGARDT19881_2} & $\sim10^{-16}$ \cite{blondel2013research3}\\
		$\tau\rightarrow e\gamma$ & $<3.3\times10^{-8}$ \cite{Aubert_2010_4} & $\sim10^{-8}-10^{-9}$ \cite{Hayasaka:2013dsa5}\\
		$\tau\rightarrow 3e$ & $<2.7\times10^{-8}$ \cite{HAYASAKA2010139_6} & $\sim10^{-9}-10^{-10}$\\
		$\tau\rightarrow \mu\gamma$ & $<4.2\times10^{-8}$ & $\sim10^{-8}-10^{-9}$\\
		$\tau\rightarrow 3\mu$ & $<2.1\times10^{-8}$ & $\sim10^{-9}-10^{-10}$\\
		\hline
	\end{tabular*}
	\caption{The upper limit of the branching ratio that is expected to be achieved by the LFV  process measurement experiment at present and in the future. }
	\label{tab1}
\end{table*}

\section{The LRSSM \label{sec2}}
The gauge group of LRSSM is $SU{{\left( 3 \right)}_{C}}\otimes SU{{\left( 2 \right)}_{L}}\otimes SU{{\left( 2 \right)}_{R}}\otimes U{{\left( 1 \right)}_{B-L}}$. Among the many types of SUSY models, LRSSM is a relatively large supersymmetric model. On the one hand, it has an additional RH interaction gauge group $SU{{\left( 2 \right)}_{R}}$ , so it will have additional $W_{R}^{\pm ,0}$  gauge bosons. At the same time, the model also has a $U{{\left( 1 \right)}_{B-L}}$ gauge group, which generates additional $Z'$ boson after the gauge symmetry is broken. On the other hand, a large number of Higgs-bosons are introduced into LRSSM, including bidoublets, triplets, and the singlet Higgs-boson. Although the LRSSM model is somewhat large, at the same time its advantages are obvious. For example, it can naturally solve the neutrino mass problem and the parity violation problem. At the same time, it disallows explicit R-parity violation, and it can provide a solution to the problem of strong and weak CP violation without introducing the axion\cite{Kuchimanchi:1993jg}. In addition, left-right symmetry is also beneficial for many extra-dimensional models and gauge unification schemes (such as SO(10)). In conclusion, LRSSM is indeed a very attractive solution for solving the current  problems in particle physics and cosmology \cite{Chatterjee_2019_9}.

The lepton and quark superfields of the LRSSM model we use are as follows
\begin{eqnarray}
	&&\hat{Q}_i=\left(\begin{array}{c}\hat u_i\\ \hat d_i\end{array}\right)=(\textbf{3}, \textbf{2}, \textbf{1}, 1/6), \quad\;\hat{Q}_i^c=\left(\begin{array}{c}\hat u_i^c\\ \hat d_i^c\end{array}\right)=(\bar{\textbf{3}}, \textbf{1}, \bar{\textbf{2}}, -1/6),\nonumber\\
	&&\nonumber \\
	&&\hat{L}_i=\left(\begin{array}{c}\hat \nu_i\\ \hat e_i\end{array}\right)=(\textbf{1}, \textbf{2}, \textbf{1}, -1/2),
	\quad\;\hat{L}_i^c=\left(\begin{array}{c}\hat \nu_i^c\\ \hat e_i^c\end{array}\right)=(\textbf{1}, \textbf{1}, \bar{\textbf{2}}, 1/2), 
\end{eqnarray}
where $i$ is the generation indicator, and at the same time the color indicators are hided. The Higss part of the model is \cite{Frank_2011_10},\cite{Basso_2015_11}
\begin{align}
	& {{\Phi }_{1}}=\left( \begin{matrix}
		H_{1}^{+} & H_{1}^{\prime 0}  \\
		H_{1}^{0} & H_{1}^{-}  \\
	\end{matrix} \right)=(\mathbf{1},\mathbf{2},\mathbf{\bar{2}},0),\quad \text{     }{{\Phi }_{2}}=\left( \begin{matrix}
		H_{2}^{+} & H_{2}^{0}  \\
		H_{2}^{\prime 0} & H_{2}^{-}  \\
	\end{matrix} \right)=(\mathbf{1},\mathbf{2},\mathbf{\bar{2}},0),\nonumber \\ 
	& \nonumber \\ 
	& \bar{\Delta }=\left( \begin{matrix}
		\frac{\delta _{1L}^{-}}{\sqrt{2}} & \delta _{1L}^{0}  \\
		\delta _{1L}^{--} & -\frac{\delta _{1L}^{-}}{\sqrt{2}}  \\
	\end{matrix} \right)=(\mathbf{1},\mathbf{3},\mathbf{1},-1),\text{   }\quad \Delta =\left( \begin{matrix}
		\frac{\delta _{2L}^{+}}{\sqrt{2}} & \delta _{2L}^{++}  \\
		\delta _{2L}^{0} & -\frac{\delta _{2L}^{+}}{\sqrt{2}}  \\
	\end{matrix} \right)=(\mathbf{1},\mathbf{3},\mathbf{1},1)\nonumber \\ 
	& \nonumber \\ 
	& {{\Delta }^{c}}=\left( \begin{matrix}
		\frac{\delta _{1R}^{-}}{\sqrt{2}} & \delta _{1R}^{0}  \\
		\delta _{1R}^{--} & -\frac{\delta _{1R}^{-}}{\sqrt{2}}  \\
	\end{matrix} \right)=(\mathbf{1},\mathbf{1},\mathbf{3},-1),\text{  }\quad {{{\bar{\Delta }}}^{c}}=\left( \begin{matrix}
		\frac{\delta _{2R}^{+}}{\sqrt{2}} & \delta _{2R}^{++}  \\
		\delta _{2R}^{0} & -\frac{\delta _{2R}^{+}}{\sqrt{2}}  \\
	\end{matrix} \right)=(\mathbf{1},\mathbf{1},\mathbf{3},1) \nonumber \\ 
	& \nonumber \\ 
	& \text{                                      }S=(\mathbf{1},\mathbf{1},\mathbf{1},0).
\end{align}
The superpotential of the model is \cite{PhysRevD.101.115014_12}
\begin{align}
	& W={{\xi }_{F}}\widehat{S}-{{\mu }_{4}}{{\widehat{\Phi }}_{1}}{{\widehat{\Phi }}_{1}}-{{\mu }_{5}}{{\widehat{\Phi }}_{2}}{{\widehat{\Phi }}_{2}}-{{Y}_{L3}}\widehat{L}\widehat{\Delta }\widehat{L}-{{Y}_{L4}}{{\widehat{L}}^{c}}{{\widehat{\Delta }}^{c}}{{\widehat{L}}^{c}}+{{Y}_{L1}}{{\widehat{L}}^{c}}\widehat{L}{{\widehat{\Phi }}_{1}}+{{Y}_{L2}}{{\widehat{L}}^{c}}\widehat{L}{{\widehat{\Phi }}_{2}} \nonumber \\ 
	& +{{Y}_{Q1}}{{\widehat{Q}}^{c}}\widehat{Q}{{\widehat{\Phi }}_{1}}+{{Y}_{Q2}}{{\widehat{Q}}^{c}}\widehat{Q}{{\widehat{\Phi }}_{2}}+{{\lambda }_{R}}\widehat{S}{{\widehat{\Delta }}^{c}}{{\widehat{\overline{\Delta }}}^{c}}+{{\lambda }_{L}}\widehat{S}\widehat{\Delta }\widehat{\overline{\Delta }}-{{\lambda }_{3}}\widehat{S}{{\widehat{\Phi }}_{1}}{{\widehat{\Phi }}_{2}}+\frac{1}{3}{{\lambda }_{S}}\widehat{S}\widehat{S}\widehat{S},
\end{align}
for the sake of brevity, we have hidden all the indicators in it. In LRSSM, the Yukawa coupling matrices need to meet certain conditions, and the left-right symmetry requires ${{Y}_{L1,2}},{{Y}_{Q1,2}}$ and ${{Y}_{L3,4}} $ symmetric \cite{PhysRevD.70.036004_13,Mohapatra:1996vg}.

The soft breaking terms of the model as follows, the part corresponding to the superpotential is
\begin{align}
	& -{{L}_{SB,W}}=2H_{1}^{-}H_{1}^{+}{{B}_{{{\mu }_{4}}}}-2H_{1}^{0}H_{1}^{\text{ }\!\!'\!\!\text{ }0}{{B}_{{{\mu }_{4}}}}+2H_{2}^{-}H_{2}^{+}{{B}_{{{\mu }_{5}}}}-2H_{2}^{0}H_{2}^{\text{ }\!\!'\!\!\text{ }0}{{B}_{{{\mu }_{5}}}}+S{{L}_{{{\xi }_{F}}}}-H_{1}^{0}H_{2}^{0}S{{T}_{{{\lambda }_{3}}}}\nonumber \\ 
	& +H_{1}^{+}H_{2}^{-}S{{T}_{{{\lambda }_{3}}}}+H_{1}^{-}H_{2}^{+}S{{T}_{{{\lambda }_{3}}}}-H_{1}^{'0}H_{2}^{'0}S{{T}_{{{\lambda }_{3}}}}+\delta _{1L}^{0}\delta _{2L}^{0}S{{T}_{{{\lambda }_{L}}}}+\delta _{1L}^{-}\delta _{2L}^{+}S{{T}_{{{\lambda }_{L}}}}+\delta _{1L}^{--}\delta _{2L}^{++}S{{T}_{{{\lambda }_{L}}}}\nonumber \\ 
	& +\delta _{1R}^{0}\delta _{2R}^{0}S{{T}_{{{\lambda }_{R}}}}+\delta _{1R}^{-}\delta _{2R}^{+}S{{T}_{{{\lambda }_{R}}}}+\delta _{1R}^{--}\delta _{2R}^{++}S{{T}_{{{\lambda }_{R}}}}+\frac{1}{3}{{S}^{3}}{{T}_{{{\lambda }_{S}}}}+H_{1}^{0}\tilde{e}_{i}^{c\text{*}}{{{\tilde{e}}}_{j}}{{T}_{L1ij}}-H_{1}^{+}\tilde{\nu }_{i}^{c\text{*}}{{{\tilde{e}}}_{j}}{{T}_{L1ij}}\nonumber \\ 
	& -H_{1}^{-}\tilde{e}_{i}^{c\text{*}}{{{\tilde{\nu }}}_{j}}{{T}_{L1ij}}+H_{1}^{'0}\tilde{\nu }_{i}^{c\text{*}}{{{\tilde{\nu }}}_{j}}{{T}_{L1ij}}+H_{2}^{'0}\tilde{e}_{i}^{c\text{*}}{{{\tilde{e}}}_{j}}{{T}_{L2ij}}-H_{2}^{+}\tilde{\nu }_{i}^{c\text{*}}{{{\tilde{e}}}_{j}}{{T}_{L2ij}}-H_{2}^{-}\tilde{e}_{i}^{c\text{*}}{{{\tilde{\nu }}}_{j}}{{T}_{L2ij}}\nonumber \\ 
	& +H_{2}^{0}\tilde{\nu }_{i}^{c\text{*}}{{{\tilde{\nu }}}_{j}}{{T}_{}}_{L2ij}+\delta _{2L}^{++}{{{\tilde{e}}}_{i}}{{{\tilde{e}}}_{k}}{{T}_{L3ik}}+\frac{1}{\sqrt{2}}\delta _{2L}^{+}{{{\tilde{e}}}_{k}}{{{\tilde{\nu }}}_{i}}{{T}_{L3ik}}+\frac{1}{\sqrt{2}}\delta _{2L}^{+}{{{\tilde{e}}}_{i}}{{{\tilde{\nu }}}_{k}}{{T}_{L3ik}}-\delta _{2L}^{0}{{{\tilde{\nu }}}_{i}}{{{\tilde{\nu }}}_{k}}{{T}_{L3ik}}\nonumber \\ 
	& -\delta _{1R}^{--}\tilde{e}_{i}^{c\text{*}}\tilde{e}_{k}^{c\text{*}}{{T}_{L4ik}}-\frac{1}{\sqrt{2}}\delta _{1R}^{-}\tilde{e}_{k}^{c\text{*}}\tilde{\nu }_{i}^{c\text{*}}{{T}_{L4ik}}-\frac{1}{\sqrt{2}}\delta _{1R}^{-}\tilde{e}_{i}^{c\text{*}}\tilde{\nu }_{k}^{c\text{*}}{{T}_{L4ik}}+\delta _{1R}^{0}\tilde{\nu }_{i}^{c\text{*}}\tilde{\nu }_{k}^{c\text{*}}{{T}_{L4ik}}\nonumber \\ 
	& +H_{1}^{0}\tilde{d}_{i\alpha }^{c\text{*}}{{\delta }_{\alpha \beta }}{{{\tilde{d}}}_{j\beta }}{{T}_{Q1ij}}-H_{1}^{+}\tilde{u}_{i\alpha }^{c\text{*}}{{\delta }_{\alpha \beta }}{{{\tilde{d}}}_{j\beta }}{{T}_{Q1ij}}-H_{1}^{-}\tilde{d}_{i\alpha }^{c\text{*}}{{\delta }_{\alpha \beta }}{{{\tilde{u}}}_{j\beta }}{{T}_{Q1ij}}+H_{1}^{\text{ }\!\!'\!\!\text{ }0}\tilde{u}_{i\alpha }^{c\text{*}}{{\delta }_{\alpha \beta }}{{{\tilde{u}}}_{j\beta }}{{T}_{Q1ij}}\nonumber \\ 
	& +H_{2}^{\text{ }\!\!'\!\!\text{ }0}\tilde{d}_{i\alpha }^{c\text{*}}{{\delta }_{\alpha \beta }}{{{\tilde{d}}}_{j\beta }}{{T}_{Q2ij}}-H_{2}^{+}\tilde{u}_{i\alpha }^{c\text{*}}{{\delta }_{\alpha \beta }}{{{\tilde{d}}}_{j\beta }}{{T}_{Q2ij}}-H_{2}^{-}\tilde{d}_{i\alpha }^{c\text{*}}{{\delta }_{\alpha \beta }}{{{\tilde{u}}}_{j\beta }}{{T}_{Q2ij}}+H_{2}^{0}\tilde{u}_{i\alpha }^{c\text{*}}{{\delta }_{\alpha \beta }}{{{\tilde{u}}}_{j\beta }}{{T}_{Q2ij}}\nonumber \\ 
	& +\text{h}.\text{c}..
\end{align}
The mass terms related to scalar fields and gauginos are
\begin{align}
	& -{{L}_{SB,\phi }}=m_{{\bar{\Delta }}}^{2}{{\left| \delta _{1L}^{0} \right|}^{2}}+m_{{\bar{\Delta }}}^{2}{{\left| \delta _{1L}^{-} \right|}^{2}}+m_{{\bar{\Delta }}}^{2}{{\left| \delta _{1L}^{--} \right|}^{2}}+m_{{{\Delta }^{c}}}^{2}{{\left| \delta _{1R}^{0} \right|}^{2}}+m_{{{\Delta }^{c}}}^{2}{{\left| \delta _{1R}^{-} \right|}^{2}}+m_{{{\Delta }^{c}}}^{2}{{\left| \delta _{1R}^{--} \right|}^{2}}\nonumber \\ 
	& +m_{\Delta }^{2}{{\left| \delta _{2L}^{0} \right|}^{2}}+m_{\Delta }^{2}{{\left| \delta _{2L}^{+} \right|}^{2}}+m_{\Delta }^{2}{{\left| \delta _{2L}^{++} \right|}^{2}}+m_{{{{\bar{\Delta }}}^{c}}}^{2}{{\left| \delta _{2R}^{0} \right|}^{2}}+m_{{{{\bar{\Delta }}}^{c}}}^{2}{{\left| \delta _{2R}^{+} \right|}^{2}}+m_{{{{\bar{\Delta }}}^{c}}}^{2}{{\left| \delta _{2R}^{++} \right|}^{2}}\nonumber \\ 
	& +m_{{{\Phi }_{1}}}^{2}{{\left| H_{1}^{0} \right|}^{2}}+m_{{{\Phi }_{1}}}^{2}{{\left| H_{1}^{-} \right|}^{2}}+m_{{{\Phi }_{1}}}^{2}{{\left| H_{1}^{+} \right|}^{2}}+m_{{{\Phi }_{1}}}^{2}{{\left| H_{1}^{\text{ }\!\!'\!\!\text{ }0} \right|}^{2}}+m_{{{\Phi }_{2}}}^{2}{{\left| H_{2}^{0} \right|}^{2}}\nonumber \\ 
	& +m_{{{\Phi }_{2}}}^{2}{{\left| H_{2}^{-} \right|}^{2}}+m_{{{\Phi }_{2}}}^{2}~{{\left| H_{2}^{+} \right|}^{2}}+m_{{{\Phi }_{2}}}^{2}{{\left| H_{2}^{'0} \right|}^{2}}+m_{S}^{2}{{\left| S \right|}^{2}}+m_{{{\Phi }_{1}}{{\Phi }_{2}}}^{2}H_{2}^{\text{ }\!\!'\!\!\text{ }0}H_{1}^{0\text{*}}\nonumber \\ 
	& +m_{{{\Phi }_{1}}{{\Phi }_{2}}}^{2}H_{2}^{-}H_{1}^{-\text{*}}+m_{{{\Phi }_{1}}{{\Phi }_{2}}}^{2}H_{2}^{+}H_{1}^{+\text{*}}+m_{{{\Phi }_{1}}{{\Phi }_{2}}}^{2}H_{2}^{0}H_{1}^{'0\text{*}}+m_{{{\Phi }_{1}}{{\Phi }_{2}}}^{2}H_{1}^{'0}H_{2}^{0\text{*}}\nonumber \\ 
	& +m_{{{\Phi }_{1}}{{\Phi }_{2}}}^{2}H_{1}^{-}H_{2}^{-\text{*}}+m_{{{\Phi }_{1}}{{\Phi }_{2}}}^{2}H_{1}^{+}H_{2}^{+\text{*}}+m_{{{\Phi }_{1}}{{\Phi }_{2}}}^{2}H_{1}^{0}H_{2}^{\text{ }\!\!'\!\!\text{ }0\text{*}}+\tilde{d}_{i\alpha }^{\text{*}}{{\delta }_{\alpha \beta }}m_{Qij}^{2}{{{\tilde{d}}}_{j\beta }}\nonumber \\ 
	& +\tilde{d}_{j\beta }^{c\text{*}}{{\delta }_{\alpha \beta }}m_{{{Q}^{c}}ij}^{2}\tilde{d}_{i\alpha }^{c}+\tilde{e}_{i}^{\text{*}}m_{Lij}^{2}{{{\tilde{e}}}_{j}}+\tilde{e}_{j}^{c\text{*}}m_{{{L}^{c}}ij}^{2}\tilde{e}_{i}^{c}+\tilde{u}_{i\alpha }^{\text{*}}{{\delta }_{\alpha \beta }}m_{Qij}^{2}{{{\tilde{u}}}_{j\beta }}\nonumber \\ 
	& +\tilde{u}_{j\beta }^{c\text{*}}{{\delta }_{\alpha \beta }}m_{{{Q}^{c}}ij}^{2}\tilde{u}_{i\alpha }^{c}+\tilde{\nu }_{i}^{\text{*}}m_{L,ij}^{2}{{{\tilde{\nu }}}_{j}}+\tilde{\nu }_{j}^{c\text{*}}m_{{{L}^{c}}ij}^{2}\tilde{\nu }_{i}^{c},  
\end{align}

\begin{align}
&-{{L}_{SB,\lambda }}=\frac{1}{2}\left( {{M}_{1}}{{{\tilde{\hat{B}}}}^{2}}+{{M}_{2L}}{{\delta }_{ij}}\tilde{W}_{L}^{i}\tilde{W}_{L}^{j}+{{M}_{2R}}{{\delta }_{ij}}\tilde{W}_{R}^{i}\tilde{W}_{R}^{j}+{{M}_{3}}{{\delta }_{ab}}{{{\tilde{g}}}^{a}}{{{\tilde{g}}}^{b}}+\text{h}.\text{c}. \right).	
\end{align}

At the time of the electroweak symmetry break, the Higgs fields acquires VEVs as \cite{Babu_2008_14},\cite{Frank:2014kma15}
\begin{align}
	& \left\langle S \right\rangle =\frac{{{v}_{S}}}{\sqrt{2}}{{e}^{i{{\alpha }_{S}}}}\text{,     }\left\langle {{\Phi }_{1}} \right\rangle =\left( \begin{matrix}
		0 & \frac{v_{1}^{'}}{\sqrt{2}}{{e}^{i{{\alpha }_{1}}}}  \\
		\frac{{{v}_{1}}}{\sqrt{2}} & 0  \\
	\end{matrix} \right),\text{     }\left\langle {{\Phi }_{2}} \right\rangle =\left( \begin{matrix}
		0 & \frac{{{v}_{2}}}{\sqrt{2}}  \\
		\frac{v_{2}^{'}}{\sqrt{2}}{{e}^{i{{\alpha }_{2}}}} & 0  \\
	\end{matrix} \right),\nonumber \\ 
	&\nonumber  \\ 
	& \left\langle {{\Delta }^{c}} \right\rangle =\left( \begin{matrix}
		0 & \frac{{{v}_{1R}}}{\sqrt{2}}  \\
		0 & 0  \\
	\end{matrix} \right)\text{,     }\left\langle {{{\bar{\Delta }}}^{c}} \right\rangle =\left( \begin{matrix}
		0 & 0  \\
		\frac{{{v}_{2R}}}{\sqrt{2}} & 0  \\
	\end{matrix} \right)\text{,     }\left\langle {\bar{\Delta }} \right\rangle =\left( \begin{matrix}
		0 & \frac{{{v}_{1L}}}{\sqrt{2}}  \\
		0 & 0  \\
	\end{matrix} \right)\text{,     }\left\langle \Delta  \right\rangle =\left( \begin{matrix}
		0 & 0  \\
		\frac{{{v}_{2L}}}{\sqrt{2}} & 0  \\
	\end{matrix} \right).  
\end{align}
Although there are many VEVs in Higgs-bosons, the relevant analysis will be greatly simplified after adopting the hierarchy of VEVs. So here we take the hierarchy
${{v}_{S}}\gg {{v}_{1R}},{{v}_{2R}}\gg {{v}_{1}},{{v}_{2}}\gg v_{1}^{'}=v_{2}^{'}={{v}_{1L}}={{v}_{2L}}\approx 0$ of VEVs, the
VEVs of $v_{1}^{'},v_{2}^{'},{{v}_{1L}},{{v}_{2L}}$ can be approximated as 0. Moreover, the phases  of VEVs satisfy:
${{\alpha }_{1}},{{\alpha }_{2}},{{\alpha }_{S}}\approx 0$. Then we have the VEVs that are simplified, and the neutral sector  of the Higgs fields can be written as
\begin{align}
	& \text{                             }S\begin{matrix}
		=\frac{1}{\sqrt{2}}{{\phi }_{S}}+\frac{1}{\sqrt{2}}{{v}_{S}}+i\frac{1}{\sqrt{2}}{{\sigma }_{S}}  \\
	\end{matrix},\text{   }\nonumber \\ 
	& H_{1}^{0}=\frac{1}{\sqrt{2}}{{\phi }_{H_{1}^{0}}}+\frac{1}{\sqrt{2}}{{v}_{1}}+i\frac{1}{\sqrt{2}}{{\sigma }_{H_{1}^{0}}}\text{,      }H_{2}^{0}=\frac{1}{\sqrt{2}}{{\phi }_{H_{2}^{0}}}+\frac{1}{\sqrt{2}}{{v}_{2}}+i\frac{1}{\sqrt{2}}{{\sigma }_{H_{2}^{0}}},\nonumber \\ 
	& \delta _{1R}^{0}=\frac{1}{\sqrt{2}}{{\phi }_{\delta _{1R}^{0}}}+\frac{1}{\sqrt{2}}{{v}_{1R}}+i\frac{1}{\sqrt{2}}{{\sigma }_{\delta _{1R}^{0}}}\text{,    }\delta _{2R}^{0}=\frac{1}{\sqrt{2}}{{\phi }_{\delta _{2R}^{0}}}+\frac{1}{\sqrt{2}}{{v}_{2R}}+i\frac{1}{\sqrt{2}}{{\sigma }_{\delta _{2R}^{0}}},\nonumber \\ 
	& H_{1}^{\text{ }\!\!'\!\!\text{ }0}=\frac{1}{\sqrt{2}}{{\phi }_{H_{1}^{\text{ }\!\!'\!\!\text{ }0}}}+i\frac{1}{\sqrt{2}}{{\sigma }_{H_{1}^{\text{ }\!\!'\!\!\text{ }0}}}\text{,                   }H_{2}^{\text{ }\!\!'\!\!\text{ }0}=\frac{1}{\sqrt{2}}{{\phi }_{H_{2}^{\text{ }\!\!'\!\!\text{ }0}}}+i\frac{1}{\sqrt{2}}{{\sigma }_{H_{2}^{\text{ }\!\!'\!\!\text{ }0}}},\text{  } \nonumber\\ 
	& \delta _{1L}^{0}=\frac{1}{\sqrt{2}}{{\phi }_{\delta _{1L}^{0}}}+i\frac{1}{\sqrt{2}}{{\sigma }_{\delta _{1L}^{0}}}\text{,                    }\delta _{2L}^{0}=\frac{1}{\sqrt{2}}{{\phi }_{\delta _{2L}^{0}}}+i\frac{1}{\sqrt{2}}{{\sigma }_{\delta _{2L}^{0}}}.  
\end{align}

 There are additional unique $W_{R\mu }^{k}$ gauge fields in the LRSSM, which become $W_{R\mu }^{\pm }$ and participate in the generation of $Z_{\mu }^{'}$  after the symmetry is broken. At the same time, the model also has an extended $U{{\left( 1 \right)}_{B-L}}$ gauge field, i.e. ${{\hat{B}}_{\mu }}$. The gauge fields  of the model are
\begin{align}
	& \text{SU}{{(3)}_{c}}\to {{V}_{3}}=\left( \mathbf{8},\mathbf{1},\mathbf{1}\text{,}0 \right)\equiv \left( {{{\tilde{g}}}^{a}},g_{\mu }^{a} \right),\nonumber \\ 
	& \text{SU}{{(2)}_{L}}\to {{V}_{2L}}=\left(\mathbf{1},\mathbf{3},\mathbf{1},0 \right)\equiv \left( \tilde{W}_{L}^{k},W_{L\mu }^{k} \right),\nonumber \\ 
	& \text{SU}{{(2)}_{R}}\to {{V}_{2R}}=\left( \mathbf{1},\mathbf{1},\mathbf{3},0 \right)\equiv \left( \tilde{W}_{R}^{k},W_{R\mu }^{k} \right),\nonumber \\ 
	& \text{U}{{(1)}_{B-L}}\to {{V}_{1}}=\left(\mathbf{1},\mathbf{1},\mathbf{1},0 \right)\equiv \left( \tilde{\hat{B}},{{{\hat{B}}}_{\mu }} \right).  
\end{align}

Unlike other models, the symmetry breaking process of LRSSM is more complex. First, the supersymmetry is broken, then the parity break, which leads to ${{g}_{L}}\ne {{g}_{R}}$. After that, the gauge group $SU{{\left( 2 \right)}_{L}}\otimes SU{{\left( 2 \right)}_{R}}\otimes U{{\left( 1 \right)}_{B-L}}$ , breaks to the $SU{{\left( 2 \right)}_{L}}\otimes U{{\left( 1 \right)}_{Y}}$ of the electroweak theory.  Finally, the $SU{{\left( 2 \right)}_{L}}\otimes U{{\left( 1 \right)}_{Y}}$ break to the $U{{\left( 1 \right)}_{em}}$ of QED.  The mixing of the latter two breaking processes and the gauge fields are shown below

1. $SU{{\left( 2 \right)}_{L}}\otimes SU{{\left( 2 \right)}_{R}}\otimes U{{\left( 1 \right)}_{B-L}}\to SU{{\left( 2 \right)}_{L}}\otimes U{{\left( 1 \right)}_{Y}}$

$W_{R\mu }^{1,2}\to W_{R\mu }^{\pm },\text{    }{{\hat{B}}_{\mu }},W_{R\mu }^{0}\to {{B}_{\mu }},Z_{\mu }^{'}$\\

2. $SU{{\left( 2 \right)}_{L}}\otimes U{{\left( 1 \right)}_{Y}}\to U{{\left( 1 \right)}_{em}}$

$W_{L\mu }^{1,2}\to W_{L\mu }^{\pm },\text{     }{{B}_{\mu }},W_{L\mu }^{0}\to {{A}_{\mu }},{{Z}_{\mu }}$
\\Then the mass matrix of the neutral gauge fields is
\begin{align}
&M_{{{V}^{0}}}^{2}=\left( \begin{matrix}
	\frac{1}{4}g_{L}^{2}\left( 4v_{L}^{2}+{{v}^{2}}+{{v}^{'2}} \right) & -\frac{1}{4}{{g}_{L}}{{g}_{R}}\left( {{v}^{2}}+{{v}^{'2}} \right) & -{{g}_{BL}}{{g}_{L}}v_{L}^{2}  \\
	-\frac{1}{4}{{g}_{L}}{{g}_{R}}\left( {{v}^{2}}+{{v}^{'2}} \right) & \frac{1}{4}g_{R}^{2}\left( 4v_{R}^{2}+{{v}^{2}}+{{v}^{'2}} \right) & -{{g}_{BL}}{{g}_{R}}v_{R}^{2}  \\
	-\hat{g}{{g}_{L}}v_{L}^{2} & -{{g}_{BL}}{{g}_{R}}v_{R}^{2} & g_{BL}^{2}\left( v_{L}^{2}+v_{R}^{2} \right)  \\
\end{matrix} \right)
\end{align}
the mass matrix of the charged gauge fields is
\begin{align}
&M_{{{V}^{\pm }}}^{2}=\left( \begin{matrix}
	\frac{1}{4}g_{L}^{2}\left( 2v_{L}^{2}+{{v}^{2}}+{{v}^{\text{ }\!\!'\!\!\text{ }2}} \right) & -\frac{1}{2}{{g}_{L}}{{g}_{R}}{{(v{v}')}^{*}}  \\
	-\frac{1}{2}{{g}_{L}}{{g}_{R}}\left( v{v}' \right) & \frac{1}{4}g_{R}^{2}\left( 2v_{R}^{2}+{{v}^{2}}+{{v}^{\text{ }\!\!'\!\!\text{ }2}} \right)  \\
\end{matrix} \right)
\end{align}
where
\begin{align}
	& v_{L}^{2}=v_{1L}^{2}+v_{2L}^{2},\text{         }v_{R}^{2}=v_{1R}^{2}+v_{2R}^{2},\nonumber \\ 
	& \begin{matrix}
		{{v}^{2}}=v_{1}^{2}+v_{2}^{2},{{v}^{'2}}=v_{1}^{'2}+v_{2}^{'2},\text{         }v{v}'={{v}_{1}}v_{1}^{'}{{e}^{i{{\alpha }_{1}}}}+{{v}_{2}}v_{2}^{'}{{e}^{i{{\alpha }_{2}}}}.  \\
	\end{matrix} 
\end{align}
 Under the hierarchical conditions of the VEVs, these matrices can be  simplified, we can obtain the parameterized hybrid form of the neutral gauge fields \cite{Alloul2013CharginoAN16}
\begin{align}
&\begin{matrix}
	\left( \begin{matrix}
		{{Z}_{\mu }}  \\
		{{A}_{\mu }}  \\
		Z_{\mu }^{'}  \\
	\end{matrix} \right)=\left( \begin{matrix}
		\text{cos}{{\theta }_{W}} & -\text{sin}{{\theta }_{W}}\text{sin}\phi  & -\text{sin}{{\theta }_{W}}\text{cos}\phi   \\
		\text{sin}{{\theta }_{W}} & \text{cos}{{\theta }_{W}}\text{sin}\phi  & \text{cos}{{\theta }_{W}}\text{cos}\phi   \\
		0 & \text{cos}\phi  & -\text{sin}\phi   \\
	\end{matrix} \right)\left( \begin{matrix}
		W_{L\mu }^{3}  \\
		W_{R\mu }^{3}  \\
		{{{\hat{B}}}_{\mu }}  \\
	\end{matrix} \right)  \\
	=\left( \begin{matrix}
		\frac{e}{{{g}_{Y}}} & -\frac{e{{g}_{Y}}}{{{g}_{L}}{{g}_{R}}} & -\frac{e{{g}_{Y}}}{{{g}_{L}}{{g}_{BL}}}  \\
		\frac{e}{{{g}_{L}}} & \frac{e}{{{g}_{R}}} & \frac{e}{{{g}_{BL}}}  \\
		0 & \frac{{{g}_{Y}}}{{{g}_{BL}}} & -\frac{{{g}_{Y}}}{{{g}_{R}}}  \\
	\end{matrix} \right)\left( \begin{matrix}
		W_{L\mu }^{3}  \\
		W_{R\mu }^{3}  \\
		{{{\hat{B}}}_{\mu }}  \\
	\end{matrix} \right)  \\
\end{matrix}
\end{align}
Then we can easily obtain the masses of the gauge fields
\begin{align}
	& m_{Z}^{2}=\frac{1}{4}\left[ g_{L}^{2}+{{\sin }^{2}}\phi g_{R}^{2} \right]{{v}^{2}}=\frac{g_{L}^{2}}{4{{\cos }^{2}}{{\theta }_{W}}}{{v}^{2}},\text{       }m_{{{Z}'}}^{2}=v_{R}^{2}\left( g_{BL}^{2}+g_{R}^{2} \right)=\frac{g_{R}^{2}}{{{\cos }^{2}}\phi }v_{R}^{2} \nonumber\\ 
	& m_{W}^{2}=\frac{g_{L}^{2}}{4}{{v}^{2}}m_{{{W}'}}^{2}=\frac{1}{2}g_{R}^{2}v_{R}^{2}. 
\end{align}
The angle of rotation can be expressed as
\begin{align}
	& \cos \phi =\frac{{{g}_{R}}}{\sqrt{g_{R}^{2}+g_{BL}^{2}}},\quad\sin \phi =\frac{{{g}_{BL}}}{\sqrt{g_{R}^{2}+g_{BL}^{2}}},\nonumber \\ 
	& \cos{{\theta }_{W}}=\frac{{{g}_{L}}}{\sqrt{g_{R}^{2}\text{si}{{\text{n}}^{2}}\phi +g_{L}^{2}}},\quad\sin{{\theta }_{W}}=\frac{{{g}_{R}}\text{sin}\phi }{\sqrt{g_{R}^{2}\text{si}{{\text{n}}^{2}}\phi +g_{L}^{2}}},  
\end{align}
these expressions will reduce the parameter space of the model. Further, we can select ${{g}_{L}}={{g}_{R}}$, and then 
\begin{align}
	& \cos {{\theta }_{W}}=\frac{{{m}_{W}}}{{{m}_{Z}}},\text{   }e=\sqrt{4\pi \alpha },\text{    }{{g}_{R}}={{g}_{L}}=\frac{e}{\sin {{\theta }_{W}}},\text{     }{{g}_{BL}}=\frac{e}{\sqrt{\text{cos}2{{\theta }_{W}}}}\text{, } v=\frac{2\text{cos}{{\theta }_{W}}{{m}_{Z}}}{{{g}_{L}}}.	
\end{align}

The Higgs part of the model is complex, and here we will perform a simple analysis of the neutral Higgs scalar mass matrices, which is essential for studying the mass of SM-like Higgs-boson. The Higgs-boson mass in the standard model is ${{m}_{h}}=125.25\pm 0.17$ GeV. In the model, the mass matrices of the neutral Higgs scalar are divided into three parts, and their bases are $\left( {{\phi }_{H_{1}^{0}}},{{\phi }_{H_{2}^{0}}},{{\phi }_{\delta _{1R}^{0}}},{{\phi }_{\delta _{2R}^{0}}},{{\phi }_{S}} \right)$,$\left( {{\phi }_{H_{1}^{'0}}},{{\phi }_{H_{2}^{'0}}} \right)$, and $\left( {{\phi }_{\delta _{1L}^{0}}},{{\phi }_{\delta _{2L}^{0}}} \right)$, the most important of which is the matrix based on the first set of scalar fields, which is \cite{PhysRevD.100.075010_17}
\begin{align}
&m_{H}^{2}=\left( \begin{matrix}
	{{m}_{{{\phi }_{H_{1}^{0}}}{{\phi }_{H_{1}^{0}}}}} & {{m}_{{{\phi }_{H_{2}^{0}}}{{\phi }_{H_{1}^{0}}}}} & {{m}_{{{\phi }_{\delta _{1R}^{0}}}{{\phi }_{H_{1}^{0}}}}} & {{m}_{{{\phi }_{\delta _{2R}^{0}}}{{\phi }_{H_{1}^{0}}}}} & {{m}_{{{\phi }_{\text{S}}}{{\phi }_{H_{1}^{0}}}}}  \\
	{{m}_{{{\phi }_{H_{1}^{0}}}{{\phi }_{H_{2}^{0}}}}} & {{m}_{{{\phi }_{H_{2}^{0}}}{{\phi }_{H_{2}^{0}}}}} & {{m}_{{{\phi }_{\delta _{1R}^{0}}}{{\phi }_{H_{2}^{0}}}}} & {{m}_{{{\phi }_{\delta _{2R}^{0}}}{{\phi }_{H_{2}^{0}}}}} & {{m}_{{{\phi }_{\text{S}}}{{\phi }_{H_{2}^{0}}}}}  \\
	{{m}_{{{\phi }_{H_{1}^{0}}}{{\phi }_{\delta _{1R}^{0}}}}} & {{m}_{{{\phi }_{H_{2}^{0}}}{{\phi }_{\delta _{1R}^{0}}}}} & {{m}_{{{\phi }_{\delta _{1R}^{0}}}{{\phi }_{\delta _{1R}^{0}}}}} & {{m}_{{{\phi }_{\delta _{2R}^{0}}}{{\phi }_{\delta _{1R}^{0}}}}} & {{m}_{{{\phi }_{\text{S}}}{{\phi }_{\delta _{1R}^{0}}}}}  \\
	{{m}_{{{\phi }_{H_{1}^{0}}}{{\phi }_{\delta _{2R}^{0}}}}} & {{m}_{{{\phi }_{H_{2}^{0}}}{{\phi }_{\delta _{2R}^{0}}}}} & {{m}_{{{\phi }_{\delta _{1R}^{0}}}{{\phi }_{\delta _{2R}^{0}}}}} & {{m}_{{{\phi }_{\delta _{2R}^{0}}}{{\phi }_{\delta _{2R}^{0}}}}} & {{m}_{{{\phi }_{\text{S}}}{{\phi }_{\delta _{2R}^{0}}}}}  \\
	{{m}_{{{\phi }_{H_{1}^{0}}}{{\phi }_{\text{S}}}}} & {{m}_{{{\phi }_{H_{2}^{0}}}{{\phi }_{\text{S}}}}} & {{m}_{{{\phi }_{\delta _{1R}^{0}}}{{\phi }_{\text{S}}}}} & {{m}_{{{\phi }_{\delta _{2R}^{0}}}{{\phi }_{\text{S}}}}} & {{m}_{{{\phi }_{\text{S}}}{{\phi }_{\text{S}}}}}  \\
\end{matrix} \right),
\end{align}
where the matrix elements are in Appendix. The mass matrices of the model can be obtained from SARAH \cite{Staub_2015_18},\cite{vicente2017computer19}.

Regarding the mass problem of the SM-like Higgs-boson, the tree mass should satisfy  the inequality as follows after considering the radiative corrections under the ${{g}_{L}}={{g}_{R}}$ condition \cite{Babu_2016_20,Zhang:2008jm}
\begin{align}
&m_{h}^{tree}\le \sqrt{2}{{m}_{W}}\simeq 113.7\text{GeV}.
\end{align}
The total mass can be easily increased to around 125 GeV with radiative corrections.

The sleptons are very important in our study, theirs mass matrix is given, which is present in base 
$\left( \tilde{e},{{{\tilde{e}}}^{c}} \right)$
\begin{align}
&m_{{\tilde{e}}}^{2}=\left( \begin{matrix}
	{{m}_{\tilde{e}{{{\tilde{e}}}^{*}}}} & m_{{{{\tilde{e}}}^{c}}{{{\tilde{e}}}^{*}}}^{\dagger }  \\
	-\frac{1}{2}{{\lambda }_{3}}{{v}_{2}}{{v}_{S}}{{Y}_{L1}}+\frac{1}{\sqrt{2}}\left( -2{{v}_{2}}{{Y}_{L2}}\mu _{5}^{*}+{{v}_{1}}{{T}_{L1}} \right) & {{m}_{{{{\tilde{e}}}^{c}}{{{\tilde{e}}}^{c*}}}}  \\
\end{matrix} \right),
\label{me}
\end{align}
where ${{m}_{\tilde{e}{{{\tilde{e}}}^{*}}}},{{m}_{{{{\tilde{e}}}^{c}}{{{\tilde{e}}}^{c*}}}}$ are  3$\times$3 matrices, which can be written as
\begin{align}
	& {{m}_{\tilde{e}{{{\tilde{e}}}^{*}}}}\begin{matrix}
		=\frac{1}{2}v_{1}^{2}Y_{L1}^{\dagger }{{Y}_{L1}}+\frac{1}{8}\mathbf{1}\left[ 2g_{BL}^{2}\left( -v_{2R}^{2}+v_{1R}^{2} \right)+g_{L}^{2}\left( -v_{1}^{2}+v_{2}^{2} \right) \right]+m_{L}^{2}\mathbf{1}  \\
	\end{matrix},\nonumber \\ 
	& {{m}_{{{{\tilde{e}}}^{c}}{{{\tilde{e}}}^{c*}}}}=\frac{1}{2}v_{1}^{2}{{Y}_{L1}}Y_{L1}^{\dagger }+\frac{1}{8}\mathbf{1}\left[ 2g_{BL}^{2}\left( -v_{1R}^{2}+v_{2R}^{2} \right)+g_{R}^{2}\left( 2v_{1R}^{2}-2v_{2R}^{2}-v_{1}^{2}+v_{2}^{2} \right) \right]+m_{{{L}^{c}}}^{2}\mathbf{1}.
	\label{mee}
\end{align}

\section{ LFV in the LRSSM\label{sec3}}
In this part, we analyze and calculate the LFV processes $l_{j}^{-}\to l_{i}^{-}\gamma$ and $l_{j}^{-}\to l_{i}^{-}l_{i}^{-}l_{i}^{+}$. Both of the two processes are calculated to the one-loop scale, and in the $l_{j}^{-}\to l_{i}^{-}l_{i}^{-}l_{i}^{+}$ process, there are also tree-level contributions.
\subsection{Rare decay $l_{j}^{-}\to l_{i}^{-}\gamma $}
\begin{figure}[H]
	\setlength{\unitlength}{1mm}
	\centering
	\includegraphics[width=5in]{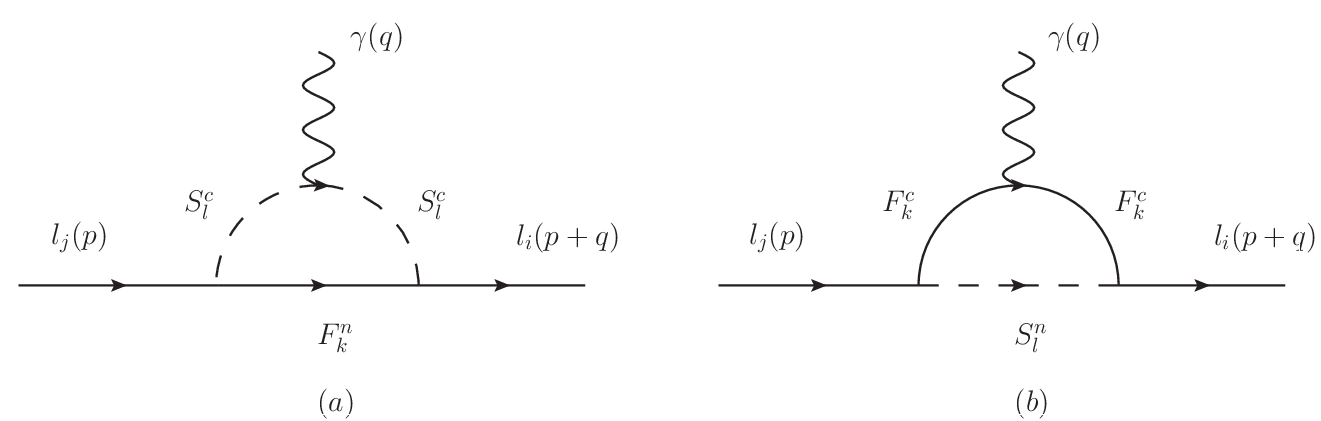}
	\vspace{0cm}
	\caption[]{ One-loop Feynman diagrams of the $l_{j}^{-}\to l_{i}^{-}\gamma $  process, where $F_{k}^{n}$ and $S_{l}^{c}$ in (a) represent neutral fermions and charged scalar particles respectively. $F_{k}^{c}$ and $S_{l}^{n}$ in (b) represent charged fermions and neutral scalar or pseudoscalar particles respectively. }
	\label{figllr}
\end{figure}

First, let's consider the Feynman diagrams of the $l_{j}^{-}\to l_{i}^{-}\gamma $ process as shown in the FIG. \ref{figllr} \cite{Yang_2020_21},\cite{Yang_2020_22}, and the off-shell invariant amplitude they contribute can be written as \cite{Yang_2021_23}
\begin{align}
&T=e{{\epsilon }^{\mu }}{{\bar{u}}_{i}}\left( p+q \right)\left[ {{q}^{2}}{{\gamma }_{\mu }}\left( A_{1}^{L}{{P}_{L}}+A_{1}^{R}{{P}_{R}} \right)+{{m}_{{{l}_{j}}}}i{{\sigma }_{\mu \nu }}{{q}^{\nu }}\left( A_{2}^{L}{{P}_{L}}+A_{2}^{R}{{P}_{R}} \right) \right]{{u}_{j}}\left( p \right),
\end{align}
where $p,q$ are the four-dimensional momentums of $l_{j}^{-},\gamma $ ,  ${{\epsilon }^{\mu }}$ is the photon polarization vector, and ${{P}_{L}},{{P}_{R}}$ are the chiral projection operators. $A_{1}^{L},A_{1}^{R},A_{2}^{L},A_{2}^{R}$ can be written as
\begin{align}
	& A_{2}^{L,R}=A_{2}^{\left( a \right)L,R}+A_{2}^{\left( b \right)L,R},\nonumber \\ 
	& A_{1}^{L,R}=A_{1}^{\left( a \right)L,R}+A_{1}^{\left( b \right)L,R},  
\end{align}
the  specific expressions of $A_{1,2}^{\left( a \right)L,R},A_{1,2}^{\left( b \right)L,R}$ are in Appendix. Then we can get the decay width of the $l_{j}^{-}\to l_{i}^{-}\gamma $ process as
\begin{align}
	 &\Gamma \left( l_{j}^{-}\to l_{i}^{-}\gamma  \right)=\frac{{{e}^{2}}}{16\pi }m_{{{l}_{j}}}^{5}\left( {{\left| A_{2}^{L} \right|}^{2}}+{{\left| A_{2}^{R} \right|}^{2}} \right).
\end{align}
Hence the branching ratio is
\begin{align}
&\text{Br}\left( l_{j}^{-}\to l_{i}^{-}\gamma  \right)=\frac{\Gamma \left( l_{j}^{-}\to l_{i}^{-}\gamma  \right)}{{{\Gamma }_{l_{j}^{-}}}},
\end{align}
where ${{\Gamma }_{l_{j}^{-}}}$ is the total decay width of the lepton $l_{j}^{-}$. We use ${{\Gamma }_{\mu }}\approx 2.996\times{{10}^{-19}}$ GeV, ${{\Gamma }_{\tau }}\approx 2.265\times{{10}^{-12}}$ GeV here \cite{S.Navas_2024_24}.
\subsection{Rare decay $l_{j}^{-}\to l_{i}^{-}l_{i}^{-}l_{i}^{+}$}
\begin{figure}[H]
	\setlength{\unitlength}{1mm}
	\centering
	\includegraphics[width=2.5in]{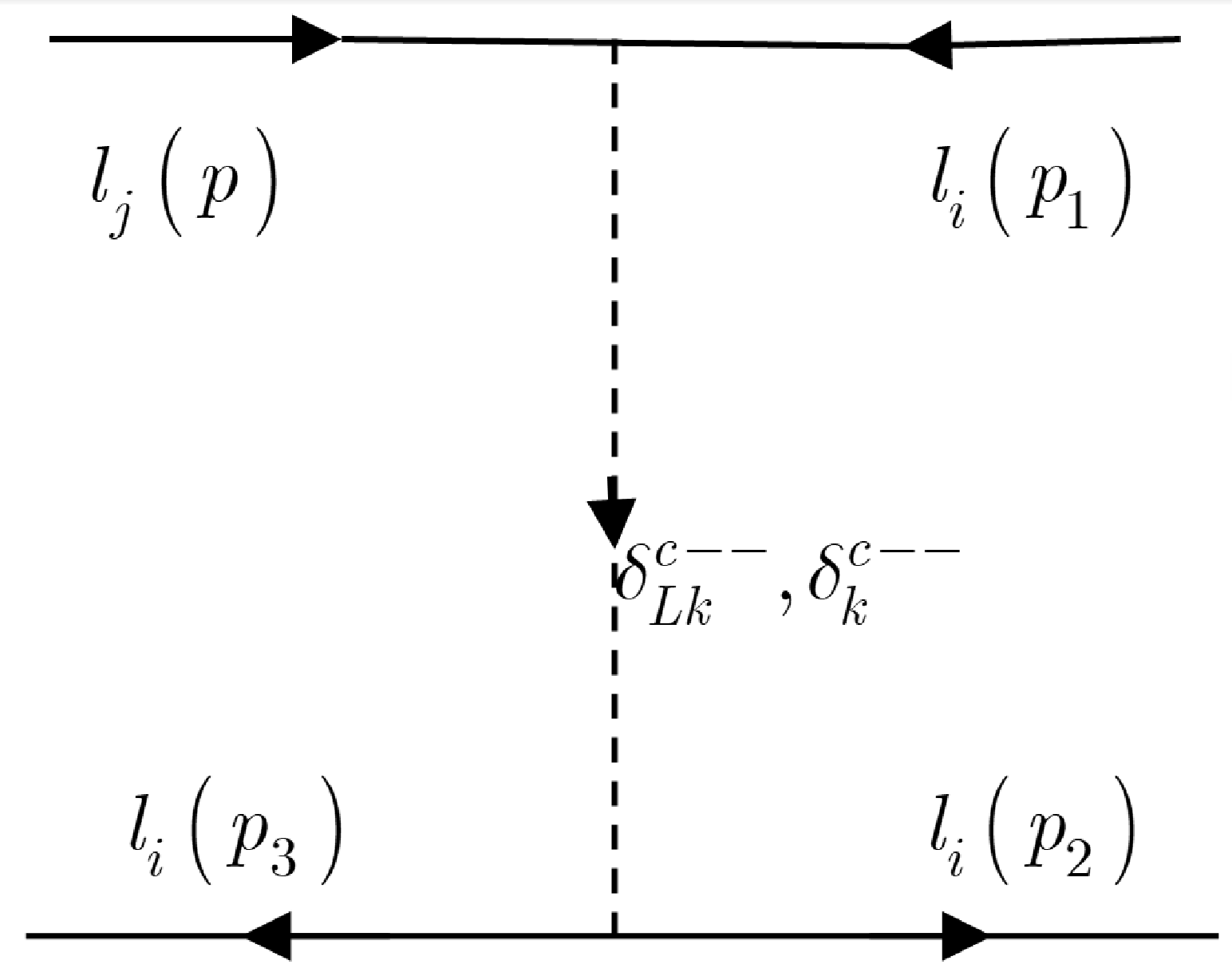}
	\vspace{0cm}
	\caption[]{The tree-level diagrams of  $l_{j}^{-}\to l_{i}^{-}l_{i}^{-}l_{i}^{+}$, where  $\delta_{Lk}^{c--}$,$\delta _{k}^{c--}$ are the double-charged Higgs-bosons in the LRSSM. }
	\label{figtree}
\end{figure}

For the $l_{j}^{-}\to l_{i}^{-}l_{i}^{-}l_{i}^{+}$ process, the leading order contributions in the LRSSM come from the tree-level diagrams mediated by the double-charged Higgs-bosons which are shown in FIG. \ref{figtree}.
The amplitude contributed from FIG. \ref{figtree} is
\begin{align}
	& T=C_{{{l}_{i}}{{l}_{j}}\delta {{_{Lk}^{c--}}^{*}}}^{L}C_{{\bar{l}}_{i}{\bar{l}}_{i}\delta _{Lk}^{c--}}^{R}\frac{i}{{{\left( p-{{p}_{1}} \right)}^{2}}-m_{Lk}^{2}}{{{\bar{u}}}_{i}}\left( {{p}_{1}} \right){{P}_{L}}{{u}_{j}}\left( p \right){{{\bar{u}}}_{i}}\left( {{p}_{3}} \right){{P}_{R}}{{v}_{i}}\left( {{p}_{2}} \right)+\nonumber \\ 
	& C_{{{l}_{i}}{{l}_{j}}\delta {{_{k}^{c--}}^{*}}}^{R}C_{{\bar{l}}_{i}{\bar{l}}_{i}\delta _{k}^{c--}}^{L}\frac{i}{{{\left( p-{{p}_{1}} \right)}^{2}}-m_{k}^{2}}{{{\bar{u}}}_{i}}\left( {{p}_{1}} \right){{P}_{R}}{{u}_{j}}\left( p \right){{{\bar{u}}}_{i}}\left( {{p}_{3}} \right){{P}_{L}}{{v}_{i}}\left( {{p}_{2}} \right), 
\end{align}
where ${{m}_{Lk}}$,${{m}_{k}}$ are the masses of the double-charged Higgs-bosons, 
the specific expressions of $C^{L,R}$ are collected in the Appendix.

\begin{figure}[H]
	\setlength{\unitlength}{1mm}
	\centering
	\includegraphics[width=2.5in]{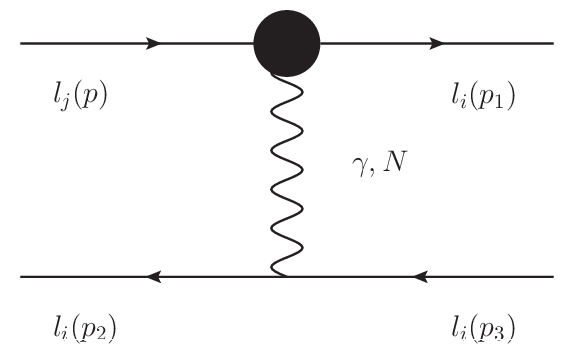}
	\vspace{0cm}
	\caption[]{The penguin-type diagrams of the $l_{j}^{-}\to l_{i}^{-}l_{i}^{-}l_{i}^{+}$ process, where $N$ represents $Z$,$Z'$ bosons, and the black dot represents the $l_{j}^{-}l_{i}^{-}\gamma $ vertex or $l_{j}^{-}l_{i}^{-}N$ vertex in FIG. \ref{figllr}. }
	\label{figl3lpg}
\end{figure}

Next, let's take a look at the penguin-type diagrams as shown in FIG. \ref{figl3lpg}. The invariant amplitude given by the penguin-type diagrams of $\gamma $ type  is
\begin{align}
	& {{T}_{\gamma -\text{penguin}}}={{{\bar{u}}}_{i}}\left( {{p}_{1}} \right)\left[ {{q}^{2}}{{\gamma }_{\mu }}\left( A_{1}^{L}{{P}_{L}}+A_{1}^{R}{{P}_{R}} \right)~+{{m}_{{{l}_{j}}}}i{{\sigma }_{\mu \nu }}{{q}^{\nu }}\left( A_{2}^{L}{{P}_{L}}+A_{2}^{R}{{P}_{R}} \right) \right]{{u}_{j}}\left( p \right) \nonumber \\ 
	&\times \frac{{{e}^{2}}}{{{q}^{2}}}{{{\bar{u}}}_{i}}\left( {{p}_{2}} \right){{\gamma }^{\mu }}{{\nu }_{i}}\left( {{p}_{3}} \right)-\left( {{p}_{1}}\leftrightarrow {{p}_{2}} \right). 
\end{align}
The  amplitude given by the penguin-type diagrams of $N$ type is
\begin{align}
	& {{T}_{N-\text{penguin}}}=\frac{{{e}^{2}}}{m_{N}^{2}}{{{\bar{u}}}_{i}}\left( {{p}_{1}} \right){{\gamma }_{\mu }}\left( {{F}^{L}}{{P}_{L}}+{{F}^{R}}{{P}_{R}} \right){{u}_{j}}\left( p \right){{{\bar{u}}}_{i}}\left( {{p}_{2}} \right) \nonumber \\ 
	&\times {{\gamma }^{\mu }}\left( C_{{{{\bar{l}}}_{i}}N{{l}_{i}}}^{L}{{P}_{L}}+C_{{{{\bar{l}}}_{i}}N{{l}_{i}}}^{R}{{P}_{R}} \right){{\nu }_{i}}\left( {{p}_{3}} \right)-\left( {{p}_{1}}\leftrightarrow {{p}_{2}} \right).
\end{align}
It can be seen that the extra ${{Z}^{'}}$ boson in the LRSSM can make contributions to the process $l_{j}^{-}\to l_{i}^{-}l_{i}^{-}l_{i}^{+}$. The specific expressions for ${{F}^{L,R}}$ are collected in the Appendix. 

\begin{figure}
	\setlength{\unitlength}{1mm}
	\centering
	\includegraphics[width=5in]{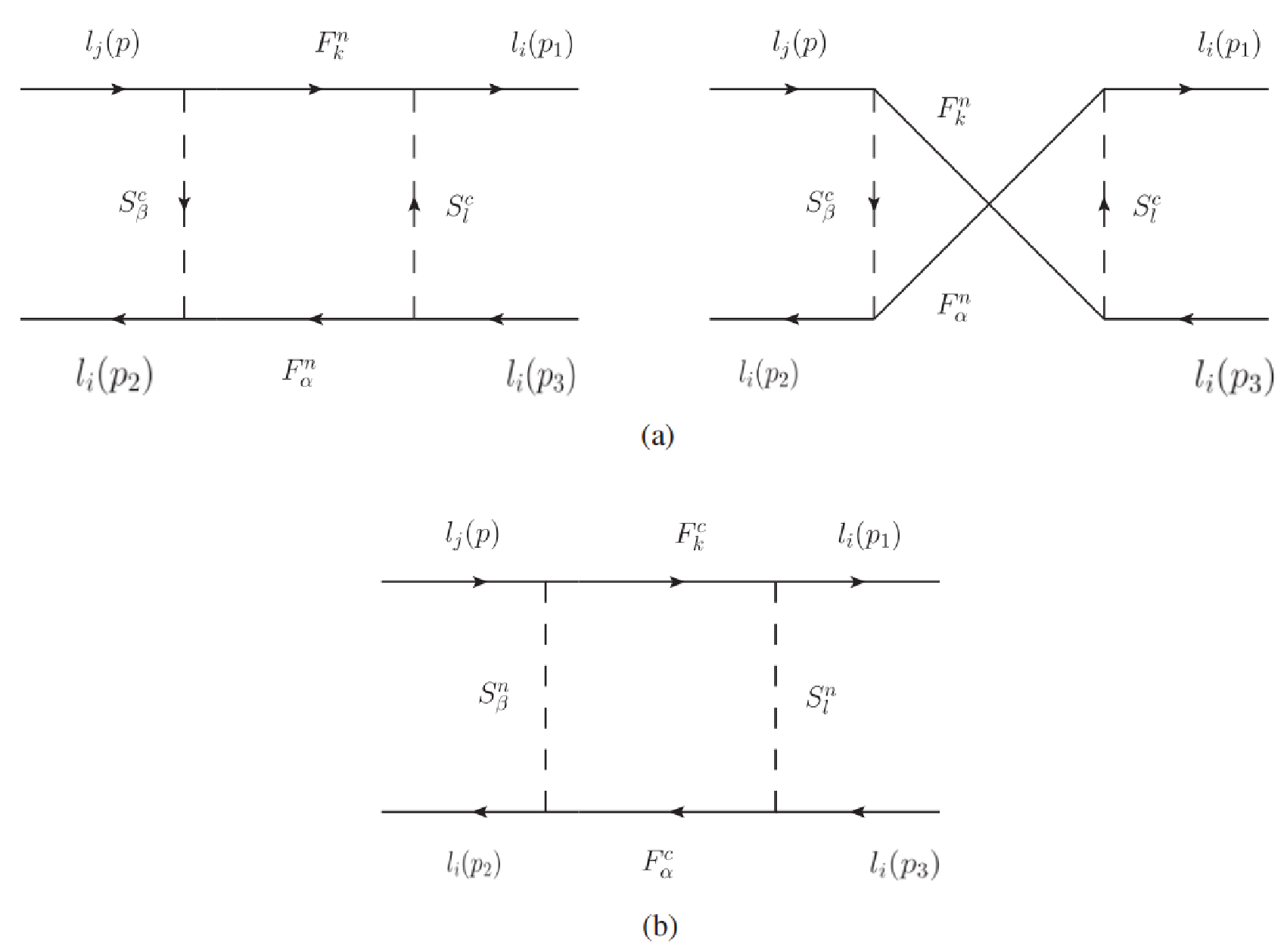}
	\vspace{0cm}
	\caption[]{box-type diagrams of the $l_{j}^{-}\to l_{i}^{-}l_{i}^{-}l_{i}^{+}$ process, where the representation of the particles is similar to that in FIG. \ref{figllr}. }
	\label{figbox}
\end{figure}

Finally, there are the box-type diagrams as shown in FIG. \ref{figbox} \cite{Yang_2019_25}.
The invariant amplitude given by these box-type diagrams is
\begin{align}
	& {{T}_{\text{box}}}=\left[ B_{1}^{L}{{e}^{2}}{{{\bar{u}}}_{i}}\left( {{p}_{1}} \right){{\gamma }_{\mu }}{{P}_{L}}{{u}_{j}}\left( p \right){{{\bar{u}}}_{i}}\left( {{p}_{2}} \right){{\gamma }^{\mu }}{{P}_{L}}{{\nu }_{i}}\left( {{p}_{3}} \right)+\left( L\leftrightarrow R \right) \right]\nonumber \\ 
	&+ \left\{ B_{2}^{L}\left[ {{e}^{2}}{{{\bar{u}}}_{i}}\left( {{p}_{1}} \right){{\gamma }_{\mu }}{{P}_{L}}{{u}_{j}}\left( p \right){{{\bar{u}}}_{i}}\left( {{p}_{2}} \right){{\gamma }^{\mu }}{{P}_{R}}{{\nu }_{i}}\left( {{p}_{3}} \right)-\left( {{p}_{1}}\leftrightarrow {{p}_{2}} \right) \right]+\left( L\leftrightarrow R \right) \right\}\nonumber \\ 
	& +\left\{ B_{3}^{L}\left[ {{e}^{2}}{{{\bar{u}}}_{i}}\left( {{p}_{1}} \right){{P}_{L}}{{u}_{j}}\left( p \right){{{\bar{u}}}_{i}}\left( {{p}_{2}} \right){{P}_{R}}{{\nu }_{i}}\left( {{p}_{3}} \right)-\left( {{p}_{1}}\leftrightarrow {{p}_{2}} \right) \right]+\left( L\leftrightarrow R \right) \right\}\nonumber \\ 
	& +\left\{ B_{4}^{L}\left[ {{e}^{2}}{{{\bar{u}}}_{i}}\left( {{p}_{1}} \right){{\sigma }_{\mu \nu }}{{P}_{L}}{{u}_{j}}\left( p \right){{{\bar{u}}}_{i}}\left( {{p}_{2}} \right){{\sigma }^{\mu \nu }}{{P}_{L}}{{\nu }_{i}}\left( {{p}_{3}} \right)-\left( {{p}_{1}}\leftrightarrow {{p}_{2}} \right) \right]+\left( L\leftrightarrow R \right) \right\} 
\end{align}
The specific expressions for $B_{1,2,3,4}^{L,R}$ are collected in the Appendix. In summary, the decay width of one-loop level can be written as \cite{Hisano_1996_26}
\begin{align}
	& \Gamma \left( l_{j}^{-}\to l_{i}^{-}l_{i}^{-}l_{i}^{+} \right)=\frac{{{e}^{4}}m_{{{l}_{j}}}^{5}}{512{{\pi }^{3}}}\Big\{ \left( {{\left| A_{2}^{L} \right|}^{2}}+{{\left| A_{2}^{R} \right|}^{2}} \right) \left( \frac{16}{3}\text{ln}\frac{{{m}_{{{l}_{j}}}}}{2{{m}_{{{l}_{i}}}}}-\frac{14}{9} \right)+\left( {{\left| A_{1}^{L} \right|}^{2}}+{{\left| A_{1}^{R} \right|}^{2}} \right)\nonumber \\ 
	&-2\left( A_{1}^{L}A_{2}^{R*}+A_{2}^{L}A_{1}^{R*}+\text{H}.\text{c}. \right)+\frac{1}{6}\left( {{\left| B_{1}^{L} \right|}^{2}}+{{\left| B_{1}^{R} \right|}^{2}} \right)+\frac{1}{3}\left( {{\left| B_{2}^{L} \right|}^{2}}+{{\left| B_{2}^{R} \right|}^{2}} \right)\nonumber \\ 
	&+\frac{1}{24}\left( {{\left| B_{3}^{L} \right|}^{2}}+{{\left| B_{3}^{R} \right|}^{2}} \right) 
	 +6\left( {{\left| B_{4}^{L} \right|}^{2}}+{{\left| B_{4}^{R} \right|}^{2}} \right)-\frac{1}{2}\left( B_{3}^{L}B_{4}^{L*}+B_{3}^{R}B_{4}^{R*}+\text{H}.\text{c}. \right)\nonumber \\ 
	&+\frac{1}{3}\left(A_{1}^{L}B_{1}^{L*}+A_{1}^{R}B_{1}^{R*}+A_{1}^{L}B_{2}^{L*}+A_{1}^{R}B_{2}^{R*}+\text{H}.\text{c}.\right)-\frac{2}{3}\left(A_{2}^{R}B_{1}^{L*}+A_{2}^{L}B_{1}^{R*}+A_{2}^{L}B_{2}^{R*}\right.\nonumber \\ 
	&+\left.A_{2}^{R}B_{2}^{R*}+\text{H}\text{.c}\text{.} \right)+\frac{1}{3}\left[ 2\left( {{\left| {{F}^{LL}} \right|}^{2}}+\right.\nonumber {{\left| {{F}^{RR}} \right|}^{2}} \right)+\left( {{\left| {{F}^{LR}} \right|}^{2}}+{{\left| {{F}^{RL}} \right|}^{2}} \right)\nonumber \\ 
	&+\left(B_{1}^{L}{{F}^{LL*}}+B_{1}^{R}{{F}^{RR*}}+B_{2}^{L}{{F}^{LR*}}+B_{2}^{R}{{F}^{RL*}}+\text{H}.\text{c}. \right)+2\left( A_{1}^{L}{{F}^{LL*}}+A_{1}^{R}{{F}^{RR*}}+\text{H}.\text{c}. \right)\nonumber \\ 
	&  +\left( A_{1}^{L}{{F}^{LR*}}+A_{1}^{R}{{F}^{RL*}}+\text{H}.\text{c}. \right)-4\left( A_{2}^{R}{{F}^{LL*}}+A_{2}^{L}{{F}^{RR*}}+\text{H}.\text{c}. \right)\nonumber \\ 
	& \left.-2\left( A_{2}^{L}{{F}^{RL*}}+A_{2}^{R}{{F}^{LR*}}+\text{H}.\text{c}. \right) \right] \Big\}, 
\end{align}
where
\begin{align}
	& {{F}^{LL}}=\underset{N=Z,{Z}'}{\mathop \sum }\,\frac{{{F}^{L}}C_{\overline{{{l}_{i}}}N{{l}_{i}}}^{L}}{m_{N}^{2}},\text{ }{{F}^{RR}}={{F}^{LL}}\left( L\leftrightarrow R \right),\text{ }\nonumber \\ 
	& {{F}^{LR}}=\underset{N=Z,{Z}'}{\mathop \sum }\,\frac{{{F}^{L}}C_{\overline{{{l}_{i}}}N{{l}_{i}}}^{R}}{m_{N}^{2}},\text{ }{{F}^{LR}}={{F}^{RL}}\left( L\leftrightarrow R \right).  
\end{align}

\section{Numerical analyses\label{sec4}}
In this part, we will perform  numerical analyses of $l_{j}^{-}\to l_{i}^{-}\gamma $ and $l_{j}^{-}\to l_{i}^{-}l_{i}^{-}l_{i}^{+}$ processes. Firstly, we choose the basic parameters ${{m}_{W}}$=80.385 GeV, ${{m}_{Z}}$=90.1876 GeV, ${{\alpha }_{em}}\left( {{m}_{Z}} \right)$=1/128.9, ${{\alpha }_{s}}\left( {{m}_{Z}} \right)$=0.118, the mass of the SM-like Higgs-boson is taken as ${{m}_{h}}=125.25\pm 0.17$ GeV.  According to the constraints of the existing research work on the parameter space of LRSSM \cite{Frank:2017tsm27},\cite{Frank_2020_28}, we choose ${{\lambda }_{L}}=0.4$, ${{\lambda }_{S}}=-0.5$, ${{T}_{{{\lambda }_{S}}}}=-2\text{ }\text{TeV}$, ${{\lambda }_{R}}=0.9$, ${{T}_{{{\lambda }_{R}}}}=-2\text{ }\text{TeV}$, ${{T}_{{{\lambda }_{3}}}}=1\text{ }\text{TeV}$, ${{M}_{3}}=3.5\text{ }\text{TeV}$, ${{\xi }_{F}}=-5000$ $\rm GeV^2$, $\text{tan}\beta =8$, $\text{tan}{{\beta }_{R}}=1.05$, $v_R=7.5\text{ }\text{TeV}$, $v_S=10\text{ } \text{TeV}$, $\lambda_3=0.10$, $M_{2L,R}=900\text{ }\text{GeV}$, where $\text{tan}\beta ={{v}_{2}}/{{v}_{1}}$, $\text{tan}{{\beta }_{R}}={{v}_{2R}}/{{v}_{1R}}$. It is worth pointing out that the above parameters are not sensitive to the results of the LFV processes. 

To simplify the numerical analysis, we can choose the slepton mass matrices $m_{L,{{L}^{c}}}^{2}=\text{diag}\left({{m}_{E}^2},{{m}_{E}^2},{{m}_{E}^2}\right)$ in the soft breaking term. As we all know, the trilinear coupling matrix ${{T}_{L1}}$ have a very drastic effect on the branching ratios of the LFV processes, so we study the variation of branching ratios with the ${{T}_{L1}}$ matrix elements of the LFV processes. ${{T}_{L1}}$ can be parameterized as
\begin{align}
&{{T}_{L1}}=\left( \begin{matrix}
	{{A}_{e}} & {{\delta }_{12}} & {{\delta }_{13}}  \\
	{{\delta }_{12}} & {{A}_{e}} & {{\delta }_{23}}  \\
	{{\delta }_{13}} & {{\delta }_{23}} & {{A}_{e}}  \\
\end{matrix} \right).
\end{align}

\begin{figure}[H]
	\setlength{\unitlength}{1mm}
	\centering
	\includegraphics[width=3.1in]{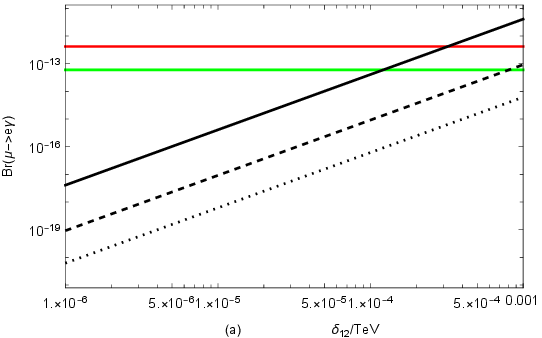}%
	\vspace{0.5cm}
	\includegraphics[width=3.1in]{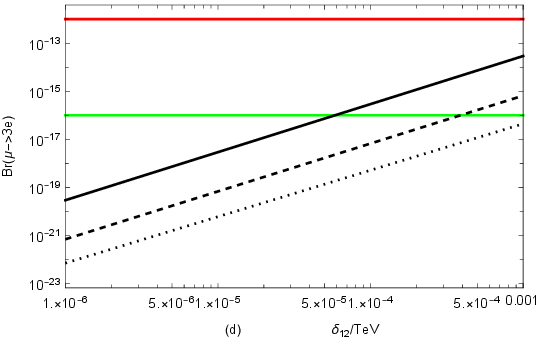}
	\vspace{0cm}
	\par
	\hspace{-0.in}
	\includegraphics[width=3.1in]{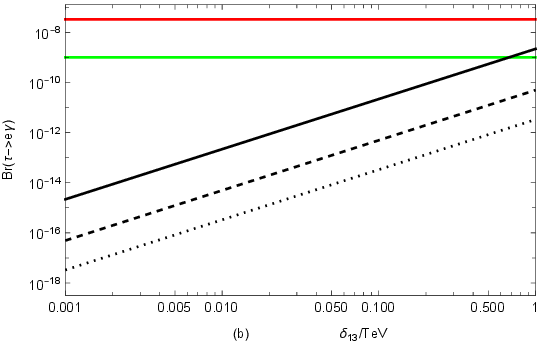}%
	\vspace{0.5cm}
	\includegraphics[width=3.1in]{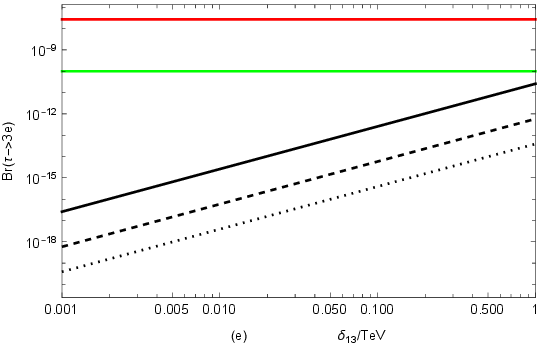}
	\vspace{0cm}
	\par
	\hspace{-0.in}
	\includegraphics[width=3.1in]{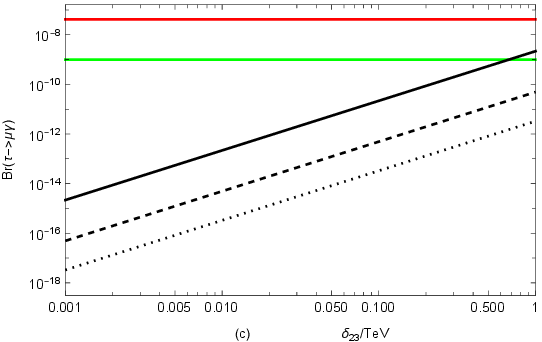}%
	\vspace{0.5cm}
	\includegraphics[width=3.1in]{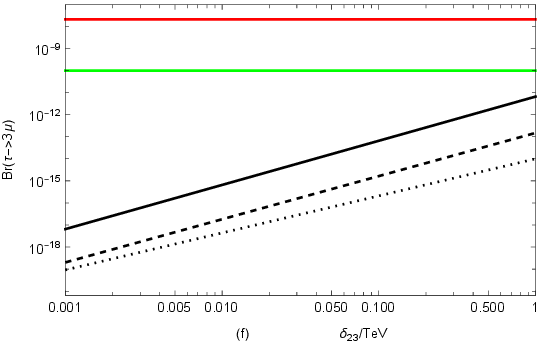}
	\vspace{0cm}
	\caption[]{Variation rules of Br($l_{j}^{-}\to l_{i}^{-}\gamma $) and Br($l_{j}^{-}\to l_{i}^{-}l_{i}^{-}l_{i}^{+}$) with matrix elements of sensitive parameter ${{T}_{L1}}$,where red lines and green lines represent present limits and future sensitivities, the solid, dashed, dotted lines correspond to ${{m}_{E}}=1.5\text{ TeV},2.5\text{ TeV},3.5\text{ TeV}$.}
	\label{6T}
\end{figure}

In order to study the variation of branching ratios in $l_{j}^{-}\to l_{i}^{-}\gamma $ and $l_{j}^{-}\to l_{i}^{-}l_{i}^{-}l_{i}^{+}$ processes with relevant sensitive parameters, it is necessary to fix some parameters in ${{T}_{L1}}$ and $m_{L,{{L}^{c}}}^{2}$ first. We choose ${{A}_{e}}$=1 TeV, and study the variation of Br($\mu \to e\gamma $), Br($\mu \to 3e$) with ${{\delta }_{12}}$ (${{\delta }_{13}}={{\delta }_{23}}=0$), the variation of Br($\tau \to e\gamma $), Br($\tau \to 3e$) with ${{\delta }_{13}}$ (${{\delta }_{12}}={{\delta }_{23}}=0$), the variation of Br($\tau \to \mu \gamma $), Br($\tau \to 3\mu $) with ${{\delta }_{23}}$ (${{\delta }_{12}}={{\delta }_{13}}=0$) under ${{m}_{E}}=1.5\text{ TeV},2.5\text{ TeV},3.5\text{ TeV}$ conditions. The images made are in FIG. \ref{6T}

From the images, it can be seen that the branching ratios of the six decay processes increase with the increase of the corresponding parameters in a certain range. As shown in  Eq (\ref{me}),(\ref{mee}), ${{T}_{L1}}$ can be influenced the mass matrix of the sleptons in turn affects the LFV process. According to present limits and future sensitivities of the decay processes, it can be seen that these datas can give certain limitations on the relevant parameters of the model. At the same time, the images of the six decay processes all follow ${{m}_{E}}$ the increase move downwards, this is also related to the Eq (\ref{me}),(\ref{mee}).

\begin{figure}[H]
	\setlength{\unitlength}{1mm}
	\centering
	\includegraphics[width=3.1in]{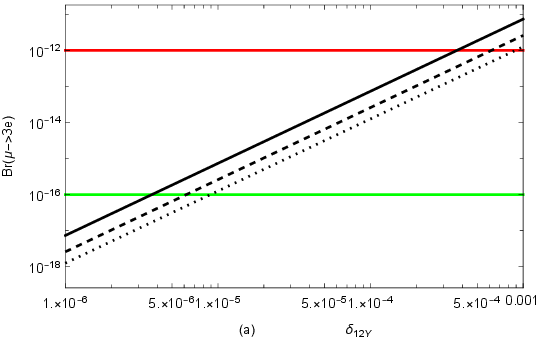}%
	\vspace{0.5cm}
	\includegraphics[width=3.1in]{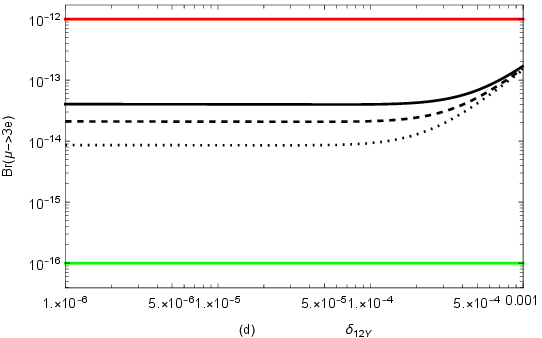}
	\vspace{0cm}
	\par
	\hspace{-0.in}
	\includegraphics[width=3.1in]{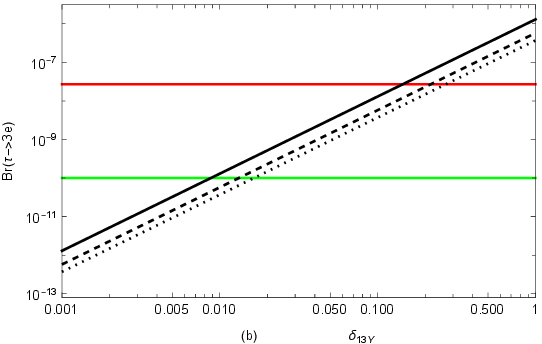}%
	\vspace{0.5cm}
	\includegraphics[width=3.1in]{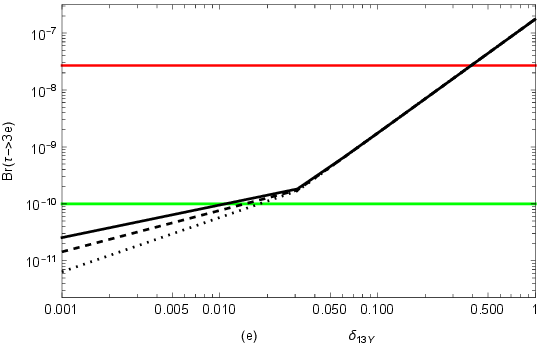}
	\vspace{0cm}
	\par
	\hspace{-0.in}
	\includegraphics[width=3.1in]{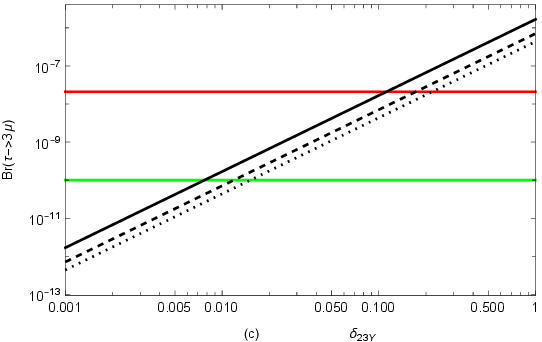}%
	\vspace{0.5cm}
	\includegraphics[width=3.1in]{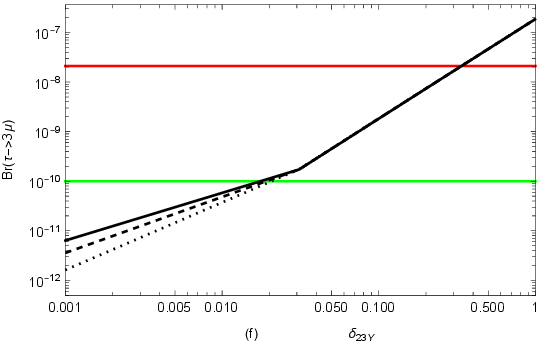}
	\vspace{0cm}
	\caption[]{Variation rules of Br($l_{j}^{-}\to l_{i}^{-}\gamma $) and Br($l_{j}^{-}\to l_{i}^{-}l_{i}^{-}l_{i}^{+}$) with matrix elements of sensitive parameter ${{Y}_{L3}}$, the solid, dashed, dotted lines correspond to ${{m}_{\Delta }}={{m}_{{\bar{\Delta }}}}=500,600,700\text{ GeV}$ in the images on the left, and the dotted, dashed, solid lines correspond to ${{\delta }_{12}}=0.5,0.75,1\text{ GeV},\text{ }{{\delta }_{13}}={{\delta }_{23}}=0.5,0.75,1\text{  TeV}$(${{m}_{\Delta }}={{m}_{{\bar{\Delta }}}}=1\text{ TeV}$) in the images on the right. The breaks of the images on the right are the results of the interference between the amplitudes of the tree-level diagrams and the one-loop diagrams. }
	\label{6Y}
\end{figure}

Analyzed in Part III.B, the tree-level contributions in the LRSSM that is specific to the $l_{j}^{-}\to l_{i}^{-}l_{i}^{-}l_{i}^{+}$ process, and the contributions are analyzed numerically in detail here. The invariant amplitude of the tree-level diagrams is sensitive to ${{Y}_{L3}}$, which is symmetrical according to the requirements of left-right symmetry, so it is possible to parameterize ${{Y}_{L3}}$ here
\begin{align}
&{{Y}_{L3}}=\left( \begin{matrix}
	0.019 & {{\delta }_{12Y}} & {{\delta }_{13Y}}  \\
	{{\delta }_{12Y}} & 0.022 & {{\delta }_{23Y}}  \\
	{{\delta }_{13Y}} & {{\delta }_{23Y}} & 0.1  \\
\end{matrix} \right).
\end{align}
Similar to the previous work, under the ${{\delta }_{12}}={{\delta }_{13}}={{\delta }_{23}}=0$ (double-charged  Higgs-bosons mass parameters ${{m}_{\Delta }}={{m}_{{\bar{\Delta }}}}=500,600,700\text{ GeV}$) and ${{\delta }_{12}}=0.5,0.75,1\text{ GeV},\text{ }{{\delta }_{13}}={{\delta }_{23}}=0.5,0.75,1\text{  TeV}$(${{m}_{\Delta }}={{m}_{{\bar{\Delta }}}}=1\text{ TeV}$)  conditions, we studied the variation of Br($\mu \to 3e$), Br($\tau \to 3e$) and Br($\tau \to 3\mu $) with respect to ${{\delta }_{12Y}}$, ${{\delta }_{13Y}}$ and ${{\delta }_{23Y}}$ respectively, as shown in FIG. \ref{6Y}.

As can be seen from the images on the left, the branching ratios of the three processes increase regularly with the increase of the corresponding parameters within a certain range. And the images move downwards as double-charged Higgs-bosons mass parameters ${{m}_{\Delta }},{{m}_{{\bar{\Delta }}}}$ increase, this shows the effect of the double-charged Higgs-boson masses on the tree-level amplitude. The images on the right are the results of the interference between the amplitudes of the tree-level diagrams and the one-loop diagrams.
And it can be seen that with the increase of the ${{Y}_{L3}}$ matrix elements ${{\delta }_{12Y}},{{\delta }_{13Y}},{{\delta }_{23Y}}$, the three curves gradually converge, which is caused by the rapid increase of the contributions of the tree-level digrams to branching ratios. The reason of these results is that ${{Y}_{L3}}$ affects the values of ${{l}_{i}}{{l}_{j}}\delta {{_{Lk}^{c--}}^{*}}$, $\bar{{{l}_{i}}}\bar{{{l}_{i}}}\delta _{Lk}^{c--}$ vertices, and the invariant amplitude of the tree-level diagrams is very dependent on these vertices. From the results of the numerical analysis, we can see that due to the 

\noindent presence of  double-charged Higgs-bosons, they can make important contributions to the $l_{j}^{-}\to l_{i}^{-}l_{i}^{-}l_{i}^{+}$ process by tree-level diagrams. 

\section{Summary\label{sec5}}
In this work, we mainly analyzed the LFV phenomena $l_{j}^{-}\to l_{i}^{-}\gamma $ and $l_{j}^{-}\to l_{i}^{-}l_{i}^{-}l_{i}^{+}$ in LRSSM, the new particles can make important contributions to the LFV processes in varying degrees compared to MSSM. In particular, the double-charged Higgs-bosons in the model make unique tree-level contributions to the $l_{j}^{-}\to l_{i}^{-}l_{i}^{-}l_{i}^{+}$ process. In terms of numerical analysis, within a certain range, Br($l_{j}^{-}\to l_{i}^{-}\gamma $) and Br($l_{j}^{-}\to l_{i}^{-}l_{i}^{-}l_{i}^{+}$) increase with the increase of the trilinear coupling matrix ${{T}_{L1}}$ matrix elements is obtained. In the  $l_{j}^{-}\to l_{i}^{-}l_{i}^{-}l_{i}^{+}$  process that contains the tree-level diagrams, the incremental variation rules of Br($l_{j}^{-}\to l_{i}^{-}l_{i}^{-}l_{i}^{+}$) with the matrix elements of the Yukawa coupling matrix  ${{Y}_{L3}}$ is concluded. In addition to this, it can be seen from the images, we limited the parameter space of the model through present limits and future sensitivities for Br($l_{j}^{-}\to l_{i}^{-}\gamma $) and Br($l_{j}^{-}\to l_{i}^{-}l_{i}^{-}l_{i}^{+}$).  In LRSSM, the predictions of the branching ratios of the LFV processes are enhanced by the new physics to an observable level, suggesting that the LFV phenomena are likely to be observed in future high-energy physics experiments.

\begin{acknowledgments}
The work is supported by the Natural Science Foundation of Guangxi Autonomous Region with Grant No. 2022GXNSFDA035068.
\end{acknowledgments}

\appendix

\section{}

The coefficients associated with FIG. \ref{figllr} are
\begin{align}
	& A_{1}^{\left( a \right)L}=\frac{1}{6m_{W}^{2}}C_{{{{\bar{l}}}_{i}}F_{k}^{n}S_{l}^{c}}^{L}C_{\bar{F}_{k}^{n}S{{_{l}^{c}}^{*}}{{l}_{j}}}^{R}{{I}_{4}}\left( {{x}_{F_{k}^{n}}},{{x}_{S_{l}^{c}}} \right),\nonumber \\ 
	& A_{2}^{\left( a \right)L}=\frac{{{m}_{F_{k}^{n}}}}{{{m}_{{{l}_{j}}}}m_{W}^{2}}C_{{{{\bar{l}}}_{i}}F_{k}^{n}S_{l}^{c}}^{L}C_{\bar{F}_{k}^{n}S_{l}^{c*}{{l}_{j}}}^{L}\left[ {{I}_{3}}\left( {{x}_{F_{k}^{n}}},{{x}_{S_{l}^{c}}} \right)-{{I}_{1}}\left( {{x}_{F_{k}^{n}}},{{x}_{S_{l}^{c}}} \right) \right],\nonumber \\ 
	& A_{1,2}^{\left( a \right)R}=A_{1,2}^{\left( a \right)L}\left( L\leftrightarrow R \right),\nonumber \\ 
	& A_{1}^{\left( b \right)L}=\frac{1}{6m_{W}^{2}}C_{{{{\bar{l}}}_{i}}F_{k}^{c}S_{l}^{n}}^{R}C_{\bar{F}_{k}^{c}S_{l}^{n}{{l}_{j}}}^{L}\left[ {{I}_{1}}\left( {{x}_{F_{k}^{c}}},{{x}_{S_{l}^{n}}} \right)-2{{I}_{2}}\left( {{x}_{F_{k}^{c}}},{{x}_{S_{l}^{n}}} \right)-{{I}_{4}}\left( {{x}_{F_{k}^{c}}},{{x}_{S_{l}^{n}}} \right) \right],\nonumber \\ 
	& A_{2}^{\left( b \right)L}=\frac{{{m}_{F_{k}^{c}}}}{{{m}_{{{l}_{j}}}}m_{W}^{2}}C_{{{{\bar{l}}}_{i}}F_{k}^{c}S_{l}^{n}}^{L}C_{\bar{F}_{k}^{c}S_{l}^{n}{{l}_{j}}}^{L}\left[ {{I}_{1}}\left( {{x}_{F_{k}^{c}}},{{x}_{S_{l}^{n}}} \right)-{{I}_{2}}\left( {{x}_{F_{k}^{c}}},{{x}_{S_{l}^{n}}} \right)-{{I}_{4}}\left( {{x}_{F_{k}^{c}}},{{x}_{S_{l}^{n}}} \right) \right],\nonumber \\ 
	& A_{1,2}^{\left( b \right)R}=A_{1,2}^{\left( b \right)L}\left( L\leftrightarrow R \right), 
\end{align}
where ${{x}_{i}}=m_{i}^{2}/m_{W}^{2}$, $C_{abc}^{L,R}$ represent the constant parts of the Feynman rules for the coupled vertices of  $a,b,c$ particles that can be got by SARAH \cite{Staub_2012_29}. The functions ${{I}_{1,2,3,4}}$ and the specific expressions  of the ${{G}_{1,2,3,4}}$ below can be found in the references \cite{Zhang:2013jva30},\cite{Zhang:2014osa31}.

The coefficients associated with the tree-level diagrams are
\begin{align}
	& C_{{{l}_{i}}{{l}_{j}}\delta {{_{Lk}^{c--}}^{*}}}^{L}=-\text{i}\left( \sum\limits_{b=1}^{3}{U_{Ljb}^{e*}\sum\limits_{a=1}^{3}{U_{Lia}^{e*}{{Y}_{L3ab}}}}+\sum\limits_{b=1}^{3}{U_{Lib}^{e*}\sum\limits_{a=1}^{3}{U_{Lja}^{e*}{{Y}_{L3ab}}}} \right)U_{Lk2}^{--} \nonumber \\ 
	& C_{{{{\bar{l}}}_{i}}{{{\bar{l}}}_{i}}\delta _{Lk}^{c--}}^{R}=-\text{i}\left( \sum\limits_{b=1}^{3}{\sum\limits_{a=1}^{3}{{{Y}_{L3ab}}U_{Lia}^{e}U_{Lib}^{e}+\sum\limits_{b=1}^{3}{\sum\limits_{a=1}^{3}{{{Y}_{L3ab}}U_{Lia}^{e}U_{Lib}^{e}}}}} \right)U_{Lk2}^{--}  \nonumber \\ 
	& C_{{{l}_{i}}{{l}_{j}}\delta {{_{k}^{c--}}^{*}}}^{R}=\text{i}\left( \sum\limits_{b=1}^{3}{\sum\limits_{a=1}^{3}{{{Y}_{L4ab}}U_{Rja}^{e}U_{Rib}^{e}+\sum\limits_{b=1}^{3}{\sum\limits_{a=1}^{3}{{{Y}_{L4ab}}U_{Ria}^{e}U_{Rjb}^{e}}}}} \right)U_{k1}^{--}  \nonumber \\ 
	& C_{{{{\bar{l}}}_{i}}{{{\bar{l}}}_{i}}\delta _{k}^{c--}}^{L}=\text{i}\left( \sum\limits_{b=1}^{3}{U_{Rib}^{e*}\sum\limits_{a=1}^{3}{U_{Ria}^{e*}{{Y}_{L4ab}}}+\sum\limits_{b=1}^{3}{U_{Rib}^{e*}\sum\limits_{a=1}^{3}{U_{Ria}^{e*}{{Y}_{L4ab}}}}} \right)U_{k1}^{--}  
\end{align}
The coefficients associated with the  penguin-type diagrams are
\begin{align}
	& {{F}^{L}}=\frac{1}{2{{e}^{2}}}C_{{{{\bar{l}}}_{i}}F_{k}^{n}S_{l}^{c}}^{R}{{C}_{S_{l}^{c*}NS_{\beta }^{c}}}C_{\bar{F}_{k}^{n}S_{\beta }^{c*}{{l}_{j}}}^{L}{{G}_{2}}\left( {{x}_{F_{k}^{n}}},{{x}_{S_{\beta }^{c}}},{{x}_{S_{l}^{c}}} \right)+\frac{{{m}_{F_{k}^{c}}}{{m}_{F_{\alpha }^{c}}}}{{{e}^{2}}m_{W}^{2}}C_{{{{\bar{l}}}_{i}}S_{\beta }^{n}F_{k}^{c}}^{R}C_{\bar{F}_{k}^{c}NF_{\alpha }^{c}}^{L}C_{\bar{F}_{\alpha }^{c}S_{\beta }^{n}{{l}_{j}}}^{L}\nonumber\\
	&{{G}_{1}}\left( {{x}_{S_{\beta }^{n}}},{{x}_{F_{k}^{c}}},{{x}_{F_{\alpha }^{c}}} \right) 
	 -\frac{1}{2{{e}^{2}}}C_{{{{\bar{l}}}_{i}}S_{\beta }^{n}F_{k}^{c}}^{R}C_{\bar{F}_{k}^{c}NF_{\alpha }^{c}}^{R}C_{\bar{F}_{\alpha }^{c}S_{\beta }^{n}{{l}_{j}}}^{L}{{G}_{2}}\left( {{x}_{S_{\beta }^{n}}},{{x}_{F_{k}^{c}}},{{x}_{F_{\alpha }^{c}}} \right),\nonumber \\ 
	& {{F}^{R}}={{F}^{L}}\left( L\leftrightarrow R \right). 
\end{align}
The coefficients associated with the box-type diagrams are
\begin{align}
	& B_{1}^{L}=\frac{{{m}_{F_{k}^{n}}}{{m}_{F_{\alpha }^{n}}}}{{{e}^{2}}m_{W}^{2}}{{G}_{3}}\left( {{x}_{F_{k}^{n}}},{{x}_{F_{\alpha }^{n}}},{{x}_{S_{\beta }^{c}}},{{x}_{S_{l}^{c}}} \right)C_{{{{\bar{l}}}_{i}}S_{l}^{c}F_{k}^{n}}^{L}C_{\bar{F}_{k}^{n}S_{\beta }^{c*}{{l}_{j}}}^{L}C_{{{{\bar{l}}}_{i}}S_{l}^{c}F_{\alpha }^{n}}^{R}C_{\bar{F}_{\alpha }^{n}S_{\beta }^{c*}{{l}_{i}}}^{R}+\frac{1}{2{{e}^{2}}m_{W}^{2}}\nonumber\\ 
	&{{G}_{4}}\left( {{x}_{F_{k}^{n}}},{{x}_{F_{\alpha }^{n}}},{{x}_{S_{\beta }^{c}}},{{x}_{S_{l}^{c}}} \right) \times \left[ C_{{{{\bar{l}}}_{i}}S_{l}^{c}F_{k}^{n}}^{R}C_{\bar{F}_{k}^{n}S_{\beta }^{c*}{{l}_{j}}}^{L}C_{{{{\bar{l}}}_{i}}S_{\beta }^{c}F_{\alpha }^{n}}^{R}C_{\bar{F}_{\alpha }^{n}S_{l}^{c*}{{l}_{i}}}^{L}+C_{{{{\bar{l}}}_{i}}S_{l}^{c}F_{k}^{n}}^{L}C_{\overline{F}_{k}^{n}S_{\beta }^{c*}{{l}_{j}}}^{R}C_{{{{\bar{l}}}_{i}}S_{l}^{c}F_{\alpha }^{n}}^{R}C_{\bar{F}_{\alpha }^{n}S_{l}^{c*}{{l}_{i}}}^{L} \right]\nonumber \\ 
	&+\frac{1}{2{{e}^{2}}m_{W}^{2}}{{G}_{4}}\left( {{x}_{F_{k}^{c}}},{{x}_{F_{\alpha }^{c}}},{{x}_{S_{\beta }^{n}}},{{x}_{S_{l}^{n}}} \right) \times C_{{{{\bar{l}}}_{i}}S_{l}^{n}F_{k}^{c}}^{R}C_{\bar{F}_{k}^{c}S_{\beta }^{n}{{l}_{j}}}^{L}C_{{{{\bar{l}}}_{i}}S_{\beta }^{n}F_{\alpha }^{c}}^{R}C_{\bar{F}_{\alpha }^{c}S_{l}^{n}{{l}_{i}}}^{L},\nonumber \\ 
	& B_{2}^{L}=-\frac{{{m}_{F_{k}^{n}}}{{m}_{F_{\alpha }^{n}}}}{2{{e}^{2}}m_{W}^{2}}{{G}_{3}}\left( {{x}_{F_{k}^{n}}},{{x}_{F_{\alpha }^{n}}},{{x}_{S_{\beta }^{c}}},{{x}_{S_{l}^{c}}} \right)C_{{{{\bar{l}}}_{i}}S_{l}^{c}F_{k}^{n}}^{R}C_{\bar{F}_{k}^{n}S_{\beta }^{c*}{{l}_{j}}}^{R}C_{{{{\bar{l}}}_{i}}S_{\beta }^{c}F_{\alpha }^{n}}^{L}C_{\bar{F}_{\alpha }^{n}S_{l}^{c*}{{l}_{i}}}^{L}+\frac{1}{4{{e}^{2}}m_{W}^{2}}\nonumber\\ 
	&{{G}_{4}}\left( {{x}_{F_{k}^{n}}},{{x}_{F_{\alpha }^{n}}},{{x}_{S_{\beta }^{c}}},{{x}_{S_{l}^{c}}} \right)  \times \left[ C_{{{{\bar{l}}}_{i}}S_{l}^{c}F_{k}^{n}}^{R}C_{\bar{F}_{k}^{n}S_{\beta }^{c*}{{l}_{j}}}^{L}C_{{{{\bar{l}}}_{i}}S_{\beta }^{c}F_{\alpha }^{n}}^{L}C_{\bar{F}_{\alpha }^{n}S_{l}^{c*}{{l}_{i}}}^{R}+C_{{{{\bar{l}}}_{i}}S_{l}^{c}F_{k}^{n}}^{R}C_{\bar{F}_{k}^{n}S_{\beta }^{c*}{{l}_{j}}}^{L}C_{{{{\bar{l}}}_{i}}S_{l}^{c}F_{\alpha }^{n}}^{R}C_{\bar{F}_{\alpha }^{n}S_{\beta }^{c*}{{l}_{i}}}^{L} \right]\nonumber\\ 
	&+\frac{1}{4{{e}^{2}}m_{W}^{2}}{{G}_{4}}\left( {{x}_{F_{k}^{c}}},{{x}_{F_{\alpha }^{c}}},{{x}_{S_{\beta }^{n}}},{{x}_{S_{l}^{n}}} \right) \times C_{{{{\bar{l}}}_{i}}S_{l}^{n}F_{k}^{c}}^{R}C_{\bar{F}_{k}^{c}S_{\beta }^{n}{{l}_{j}}}^{L}C_{{{{\bar{l}}}_{i}}S_{\beta }^{n}F_{\alpha }^{c}}^{L}C_{\bar{F}_{\alpha }^{c}S_{l}^{n}{{l}_{i}}}^{R}-\frac{{{m}_{F_{k}^{c}}}{{m}_{F_{\alpha }^{c}}}}{2{{e}^{2}}m_{W}^{2}}\nonumber\\ 
	&{{G}_{3}}\left( {{x}_{F_{k}^{c}}},{{x}_{F_{\alpha }^{c}}},{{x}_{S_{\beta }^{n}}},{{x}_{S_{l}^{n}}} \right)\times C_{{{{\bar{l}}}_{i}}S_{l}^{n}F_{k}^{c}}^{R}C_{\bar{F}_{k}^{c}S_{\beta }^{n}{{l}_{j}}}^{R}C_{{{{\bar{l}}}_{i}}S_{\beta }^{n}F_{\alpha }^{c}}^{L}C_{\bar{F}_{\alpha }^{c}S_{l}^{n}{{l}_{i}}}^{L},\nonumber \\ 
	& B_{3}^{L}=\frac{{{m}_{F_{k}^{n}}}{{m}_{F_{\alpha }^{n}}}}{{{e}^{2}}m_{W}^{2}}{{G}_{3}}\left( {{x}_{F_{k}^{n}}},{{x}_{F_{\alpha }^{n}}},{{x}_{S_{\beta }^{c}}},{{x}_{S_{l}^{c}}} \right)\left[ C_{{{{\bar{l}}}_{i}}S_{l}^{c}F_{k}^{n}}^{L}C_{\bar{F}_{k}^{n}S{{_{\beta }^{c}}^{*}}{{l}_{j}}}^{L}C_{{{{\bar{l}}}_{i}}S_{\beta }^{c}F_{\alpha }^{n}}^{L}C_{\bar{F}_{\alpha }^{n}S{{_{l}^{c}}^{*}}{{l}_{i}}}^{L}-\frac{1}{2}C_{{{{\bar{l}}}_{i}}S_{l}^{c}F_{k}^{n}}^{L}C_{\bar{F}_{k}^{n}S{{_{\beta }^{c}}^{*}}{{l}_{j}}}^{L}\right.\nonumber\\ 
	&\left. C_{{{{\bar{l}}}_{i}}S_{l}^{c}F_{\alpha }^{n}}^{L}C_{\bar{F}_{\alpha }^{n}S{{_{\beta }^{c}}^{*}}{{l}_{i}}}^{L} \right]+\frac{{{m}_{F_{k}^{c}}}{{m}_{F_{\alpha }^{c}}}}{2{{e}^{2}}m_{W}^{2}}{{G}_{3}}\left( {{x}_{F_{k}^{c}}},{{x}_{F_{\alpha }^{c}}},{{x}_{S_{\beta }^{n}}},{{x}_{S_{l}^{n}}} \right)\times C_{{{{\bar{l}}}_{i}}S_{l}^{n}F_{k}^{c}}^{R}C_{\bar{F}_{k}^{c}S_{\beta }^{n}{{l}_{j}}}^{R}C_{{{{\bar{l}}}_{i}}S_{\beta }^{n}F_{\alpha }^{c}}^{L}C_{\bar{F}_{\alpha }^{c}S_{l}^{n}{{l}_{i}}}^{L},\nonumber \\ 
	& B_{4}^{L}=\frac{{{m}_{F_{k}^{n}}}{{m}_{F_{\alpha }^{n}}}}{8{{e}^{2}}m_{W}^{2}}{{G}_{3}}\left( {{x}_{F_{k}^{n}}},{{x}_{F_{\alpha }^{n}}},{{x}_{S_{\beta }^{c}}},{{x}_{S_{l}^{c}}} \right)C_{{{{\bar{l}}}_{i}}S_{l}^{c}F_{k}^{n}}^{L}C_{\bar{F}_{k}^{n}S_{\beta }^{c*}{{l}_{j}}}^{L}C_{{{{\bar{l}}}_{i}}S_{l}^{c}F_{\alpha }^{n}}^{L}C_{\bar{F}_{\alpha }^{n}S_{\beta }^{c*}{{l}_{i}}}^{L},\nonumber \\ 
	& B_{1,2,3,4}^{R}=B_{1,2,3,4}^{L}\left( L\leftrightarrow R \right). 
\end{align}

The matrix elements of the neutral Higgs-bosons are
\begin{align}
	& {{m}_{{{\phi }_{H_{1}^{0}}}{{\phi }_{H_{1}^{0}}}}}=4|{{\mu }_{4}}{{|}^{2}}+\frac{1}{2}\lambda _{3}^{2}\left( v_{2}^{2}+v_{S}^{2} \right)+\frac{1}{8}\left[ g_{L}^{2}\left( 3v_{1}^{2}-v_{2}^{2} \right)+g_{R}^{2}\left( -2v_{1R}^{2}+2v_{2R}^{2}+3v_{1}^{2}-v_{2}^{2} \right) \right]\nonumber \\ 
	& +m_{{{\Phi }_{1}}}^{2}\nonumber \\ 
	& {{m}_{{{\phi }_{H_{1}^{0}}}{{\phi }_{H_{2}^{0}}}}}=\frac{1}{2}\left\{ {{\lambda }_{3}}\left[ 2{{\lambda }_{3}}{{v}_{1}}{{v}_{2}}-2\Re \left( {{\xi }_{F}} \right)-{{\lambda }_{R}}{{v}_{1R}}{{v}_{2R}}-{{\lambda }_{S}}v_{S}^{2} \right]-\sqrt{2}{{v}_{S}}{{T}_{{{\lambda }_{3}}}} \right\}-\frac{1}{4}\left( g_{L}^{2}+g_{R}^{2} \right){{v}_{1}}{{v}_{2}} \nonumber\\ 
	& {{m}_{{{\phi }_{H_{2}^{0}}}{{\phi }_{H_{2}^{0}}}}}=4|{{\mu }_{5}}{{|}^{2}}+\frac{1}{2}\lambda _{3}^{2}\left( v_{1}^{2}+v_{S}^{2} \right)+\frac{1}{8}\left[ -g_{L}^{2}\left( -3v_{2}^{2}+v_{1}^{2} \right)-g_{R}^{2}\left( -2v_{1R}^{2}+2v_{2R}^{2}-3v_{2}^{2}+v_{1}^{2} \right) \right]\nonumber \\ 
	& +m_{{{\Phi }_{2}}}^{2}\nonumber \\ 
	& {{m}_{{{\phi }_{H_{1}^{0}}}{{\phi }_{\delta _{1R}^{0}}}}}=-\frac{1}{2}g_{R}^{2}{{v}_{1}}{{v}_{1R}}-\frac{1}{2}{{\lambda }_{3}}{{\lambda }_{R}}{{v}_{2}}{{v}_{2R}}\nonumber \\ 
	& {{m}_{{{\phi }_{H_{2}^{0}}}{{\phi }_{\delta _{1R}^{0}}}}}=\frac{1}{2}g_{R}^{2}{{v}_{1R}}{{v}_{2}}-\frac{1}{2}{{\lambda }_{3}}{{\lambda }_{R}}{{v}_{1}}{{v}_{2R}}
\end{align}

\begin{align}
	& {{m}_{{{\phi }_{\delta _{1R}^{0}}}{{\phi }_{\delta _{1R}^{0}}}}}=\frac{1}{2}\lambda _{R}^{2}\left( v_{2R}^{2}+v_{S}^{2} \right)+\frac{1}{4}\left[ 2g_{BL}^{2}\left( 3v_{1R}^{2}-v_{2R}^{2} \right)+g_{R}^{2}\left( -2v_{2R}^{2}+6v_{1R}^{2}-v_{1}^{2}+v_{2}^{2} \right) \right]+m_{{{\Delta }^{c}}}^{2}\nonumber \\ 
	& {{m}_{{{\phi }_{H_{1}^{0}}}{{\phi }_{\delta _{2R}^{0}}}}}=\frac{1}{2}g_{R}^{2}{{v}_{1}}{{v}_{2R}}-\frac{1}{2}{{\lambda }_{3}}{{\lambda }_{R}}{{v}_{1R}}{{v}_{2}}\nonumber \\ 
	& {{m}_{{{\phi }_{H_{2}^{0}}}{{\phi }_{\delta _{2R}^{0}}}}}=-\frac{1}{2}g_{R}^{2}{{v}_{2}}{{v}_{2R}}-\frac{1}{2}{{\lambda }_{3}}{{\lambda }_{R}}{{v}_{1}}{{v}_{1R}}\nonumber \\ 
	& {{m}_{{{\phi }_{\delta _{1R}^{0}}}{{\phi }_{\delta _{2R}^{0}}}}}=\frac{1}{\sqrt{2}}{{v}_{S}}{{T}_{{{\lambda }_{R}}}}-\left( g_{BL}^{2}+g_{R}^{2} \right){{v}_{1R}}{{v}_{2R}}+{{\lambda }_{R}}\left[ \frac{1}{2}\left( 2{{\lambda }_{R}}{{v}_{1R}}{{v}_{2R}}-{{\lambda }_{3}}{{v}_{1}}{{v}_{2}}+{{\lambda }_{S}}v_{S}^{2} \right)+\Re \left( {{\xi }_{F}} \right) \right]\nonumber \\ 
	& {{m}_{{{\phi }_{\delta _{2R}^{0}}}{{\phi }_{\delta _{2R}^{0}}}}}=\frac{1}{2}\lambda _{R}^{2}\left( v_{1R}^{2}+v_{S}^{2} \right)+\frac{1}{4}\left[ -2g_{BL}^{2}\left( -3v_{2R}^{2}+v_{1R}^{2} \right)+g_{R}^{2}\left( -2v_{1R}^{2}+6v_{2R}^{2}-v_{2}^{2}+v_{1}^{2} \right) \right]\nonumber \\
	&+m_{{{\overline{\Delta }}^{c}}}^{2}\nonumber \\ 
	& {{m}_{{{\phi }_{H_{1}^{0}}}{{\phi }_{S}}}}=-\frac{1}{\sqrt{2}}{{v}_{2}}{{T}_{{{\lambda }_{3}}}}+{{\lambda }_{3}}\left( {{\lambda }_{3}}{{v}_{1}}-{{\lambda }_{S}}{{v}_{2}} \right){{v}_{S}}\nonumber \\ 
	& {{m}_{{{\phi }_{H_{2}^{0}}}{{\phi }_{S}}}}=-\frac{1}{\sqrt{2}}{{v}_{1}}{{T}_{{{\lambda }_{3}}}}+{{\lambda }_{3}}\left( {{\lambda }_{3}}{{v}_{2}}-{{\lambda }_{S}}{{v}_{1}} \right){{v}_{S}}\nonumber \\ 
	& {{m}_{{{\phi }_{\delta _{1R}^{0}}}{{\phi }_{S}}}}=\frac{1}{\sqrt{2}}{{v}_{2R}}{{T}_{{{\lambda }_{R}}}}+{{\lambda }_{R}}\left( {{\lambda }_{R}}{{v}_{1R}}+{{\lambda }_{S}}{{v}_{2R}} \right){{v}_{S}}\nonumber \\ 
	& {{m}_{{{\phi }_{\delta _{2R}^{0}}}{{\phi }_{S}}}}=\frac{1}{\sqrt{2}}{{v}_{1R}}{{T}_{{{\lambda }_{R}}}}+{{\lambda }_{R}}\left( {{\lambda }_{R}}{{v}_{2R}}+{{\lambda }_{S}}{{v}_{1R}} \right){{v}_{S}}\nonumber \\ 
	& {{m}_{{{\phi }_{S}}{{\phi }_{S}}}}=\frac{1}{2}\left\{ 2\sqrt{2}{{v}_{S}}{{T}_{{{\lambda }_{S}}}}+6\lambda _{S}^{2}v_{S}^{2}+\lambda _{3}^{2}\left( v_{1}^{2}+v_{2}^{2} \right)+{{\lambda }_{R}}\left[ 2{{\lambda }_{S}}{{v}_{1R}}{{v}_{2R}}+{{\lambda }_{R}}\left( v_{1R}^{2}+v_{2R}^{2} \right) \right]+ \right.\nonumber \\ 
	& \left. {{\lambda }_{S}}\left[ -2{{\lambda }_{3}}{{v}_{1}}{{v}_{2}}+4\Re \left( {{\xi }_{F}} \right) \right] \right\}+m_{S}^{2}. 
\end{align}

\bibliographystyle{unsrt}

\end{document}